\documentclass[preprints,review,accept,oneauthor,12pt]{Definitions/mdpi} 

\firstpage{1} 
\pubvolume{0}
\issuenum{0}
\articlenumber{0}
\pubyear{2020}
\copyrightyear{2020}
\history{Received: 2019/12/20; Accepted: 2020/01/25; Published: 2020/02/05}

\usepackage{graphicx}
\usepackage{amsmath}
\usepackage{verbatim}
\usepackage{multirow}

 \Title{Radiation-Driven Stellar Eruptions    
    \footnote{{\normalsize This is a slightly improved version of an  
   article in the e-book {\it Luminous Stars in Nearby Galaxies,}  \\    
   {\tt https://www.mdpi.com/journal/galaxies/special{\_}issues/Luminous} 
   \ (= {\it Galaxies\/} {\bf 2020}, {\it 8,} 10).  } } }    

\Author{Kris Davidson} 

\address[1]{Minnesota Institute for Astrophysics, University of 
Minnesota, 116 Church St. SE, Minneapolis, MN 55455, USA; kd@umn.edu; Tel.: +1-612-624-5711}  

\abstract{ Very massive stars occasionally expel material in colossal eruptions, 
driven by continuum radiation pressure rather than blast waves.  Some of them 
rival supernovae in total radiative output, and the mass loss is crucial for  
subsequent evolution.  Some are supernova impostors, including SN precursor 
outbursts, while others are true SN events shrouded by material that was 
ejected earlier.  Luminous Blue Variable stars (LBV's) are traditionally 
cited in relation with giant eruptions, though this connection is not well 
established.   After four decades of research,  {\it the fundamental causes 
of giant eruptions and LBV events remain elusive.}  This review outlines     
the basic relevant physics, with a brief summary of essential  
observational facts.  Reasons are described for the spectrum and emergent 
radiation temperature of an opaque outflow.  Many proposed mechanisms 
are noted for instabilities 
in the star's photosphere, in its iron opacity peak zones, and 
in its central region.  Some of the remarks and conjectures here have not 
yet become familiar in the published literature.
}  


\begin{document}


\def \resc{r_\mathrm{esc}}
\def \Tph{T_\mathrm{ph}}  
\def \sp{_\mathrm{ph}} 
\def \tautot{\tau_\mathrm{tot}}  
\def \tauth{\tau_\mathrm{th}} 
\def \tauabs{\tau_\mathrm{abs}} 
\def \tausc{\tau_\mathrm{sc}}  
\def \kabs{\kappa_\mathrm{abs}} 
\def \ksc{\kappa_\mathrm{sc}}
\def \ktot{\kappa_\mathrm{tot}} 
\def \kth{\kappa_\mathrm{th}} 
\def \avtausc{\langle\tau_\mathrm{sc}\rangle_\mathrm{em}} 
\def \teff{T_\mathrm{eff}} 


   \vspace{1mm}  

\noindent  {\bf Contents} 
\\ 1. The topic and why it matters  
\\ 2. History, 1965--2020  
\\ 3. Categories: giant eruptions, SN impostors and precursors, LBV's   
\\ 4. Apparent temperature and spectrum of a massive outflow  
\\ 5. Culpable instabilities and locales within a very massive star  
\\ 6. Other issues, especially re.\ the most-observed example     

   \vspace{2mm}  

\section{Super-Eddington Events in Massive~Stars}\label{1}  

Very massive stars lose much---and possibly most---of their 
mass in sporadic events driven by continuum radiation.  This fact has 
dire consequences for any attempt to predict the star's evolution.  
After four decades of research,  the~instability mechanism has not 
yet been established;  it may 
occur in the stellar core, or else in a subsurface  locale,  or~
conceivably at the base of the photosphere.   Without concrete models 
of this process, massive-star evolution codes can generate only 
``proof of concept'' simulations, not predictive models, because~they 
rely on assumed mass-loss rates adjusted to give plausible results. 
Even worse, eruptions may illustrate the butterfly effect: 
the time and strength   
of each outburst may depend on seemingly minor details, and~the 
total mass loss may differ greatly between two stars that appear 
identical at birth.  And~an unexpected sub-topic, involving precursors 
to  supernova events, arose about ten years ago.  Altogether, the~
most luminous stars cannot be understood without a greatly improved 
theory of radiative mass-loss events.  No theorist predicted 
any of the main observational discoveries in this~subject.      

   \vspace{1mm}
  
Most of the phenomena explored here are either giant eruptions 
(including supernova impostors, supernova precursors, and~shrouded   
supernovae) or LBV outbursts.  
They have four attributes in~common:  
\begin{itemize} 
\item Their $L/M$ ratios are near or above the Eddington Limit.  
\item Outflow speeds are usually between 100 and 800 km s$^{-1}$, 
much slower than SN blast waves. 
\item The eruptive photosphere temperatures range from 6000 to 20,000 K, 
providing enough free electrons for substantial opacity.  
\item Observed durations are much longer than relevant dynamical timescales.  
\end{itemize}    

Giant eruptions are presumably driven by continuum radiation. 
They carry far too much kinetic energy to be 
``line-driven winds.'' 
Gas pressure is quite inadequate, and blast waves are either absent or 
inconspicuous.  MHD processes, bulk turbulent pressure, and rotational
effects may play appreciable roles, 
but probably cannot exceed radiation pressure in these   
very luminous objects.  Individual events have    
peak luminosities ranging from $10^5 \, L_\odot$ to more than 
$10^8 \, L_\odot$ while ejecting masses ranging from 
$10^{-3} \, M_\odot$ to $10 \, M_\odot$ or more.     
Note that continuum radiation (i.e., a~super-Eddington 
flow) is not in itself the ``cause'' of an eruption.  Logically the 
root cause must be some process or instability that either increases 
the local radiation flux, or enhances its ability to push a 
mass~outflow. 

  \vspace{1mm} 

This review does not include eruptions with $L < 10^{5.5} \, L_\odot$, 
such as ``red transients'' and nova-like displays.  Such events generally   
involve stars with $M < 20 \, M_\odot$ or even $M < 10 \, M_\odot$, 
which are vastly more numerous than the very massive stars that  
produce giant eruptions ($M_\mathrm{ZAMS} > 50 \, M_\odot$).     
Most likely they are highly abnormal 
phenomena (e.g.,\ stellar mergers) that occur in only a tiny fraction 
of the stars.  Giant eruptions, by~contrast, tend to resemble 
each other regardless of their causative instabilities, and~ 
probably occur in a substantial fraction of the most massive stars 
-- perhaps even most of them.  
Much of Section~\ref{4} and part of Section~\ref{5} may apply also to the lower-luminosity 
eruptions, however.   

  \vspace{1mm}  
  
This article is a descriptive review like a textbook chapter, not 
a survey of publications.  It outlines the basic physics and theoretical
results with only a minimal account of the observational data.   
It also includes comments about some of the quoted results, with~a few  
personal conjectures.  Some of the generalities  
sketched here have been unfamiliar to most astronomers, even those 
who work on supernovae.  They are conceptually simple if we 
refrain from exploring technicalities.  One important topic---rotation---
is mostly neglected here, because it would  
greatly lengthen the narrative and there is not yet any strong 
evidence that it is required for the chief processes.     
Binary systems are also neglected,  except~the special case 
of $\eta$ Car;  see remarks in Section~\ref{6}.   

   \vspace{4mm}  

\section{A Checkered~History} \label{2} 

   \vspace{2mm}  

An appendix in ref.\ \cite{hd94} summarized the origins of this topic.
Since only three or four giant eruptions and supernova impostors were 
known before 2000,  they were conflated with LBV 
outbursts.  Early discoveries followed two paths, and~today we're 
not yet sure whether those paths really merge.  
First, the~examples of $\eta$ Car, P Cyg, SN 1961v, and~SN 1954J were 
known before 1970~\cite{zw65}.  Their outbursts produced supernova-like 
amounts of radiation,  but~with longer durations  than a supernova 
and the stars survived.\footnote{
   SN 1961v may have been a true supernova \cite{ko11,vdm12}, but, 
   ironically, that doesn't alter its historical role.} 
The second path began with the recognition 
in 1979 of an upper boundary in the empirical HR diagram, 
the diagonal line in the middle of Figure~1~\cite{hd79}.  Almost 
no stars are found above and to the right of that line, and~the rare
exceptions are temporary.  Since massive stars 
evolve almost horizontally across the diagram, this boundary indicates 
some sort of barrier to the outer-layer evolution of stars with 
$M > 50 \; M_\odot$.  The~probable explanation involves episodic mass 
loss as~follows.  

  \vspace{1mm}

\begin{figure}[H]   
  \centering
  \includegraphics[width=11cm]{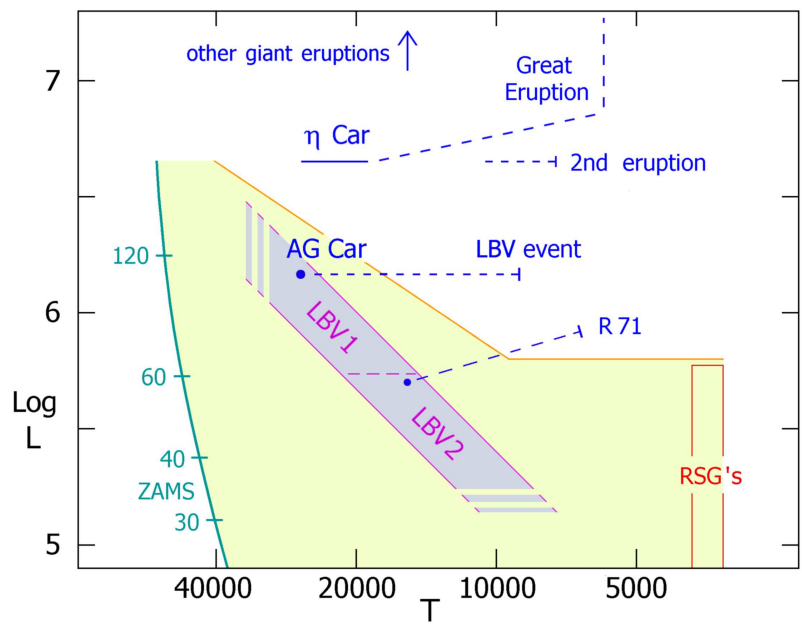}
  \caption{The empirical upper boundary and LBV instability strip in the 
   Hertzsprung-Russell Diagram.  In~reality they are ill-defined and may 
   depend on rotation and chemical composition.  The~interval between 
   the LBV strip and the boundary is very uncertain.  The~zero-age main 
   sequence on the left side shows initial masses, and~most of a very 
   massive star's evolution occurs at roughly twice the initial luminosity. }\label{f1}
  \end{figure}

Note the various zones and boundaries in Figure~\ref{f1}, though~in reality 
they are not so well defined.  If~a massive star loses a 
considerable fraction of its mass, then it cannot evolve far toward the 
right in the HR diagram.  Thus a good way to explain the HRD boundary 
is to suppose that stars above 50 $M_\odot$ lose mass in some process 
that exceeds their line-driven winds.   The~S Doradus class 
of variable stars occurs in the ``LBV1'' and ``LBV2'' zones in Figure~\ref{f1}, 
to the left of the empirical boundary. S Dor stars are remarkably close to the 
Eddington Limit (Section~\ref{3.1} below), and~they exhibit sporadic outbursts 
which can expel more material than their normal winds.  If~every 
star above 50 $M_\odot$ behaves in that way after it evolves into 
the LBV1 zone, then the boundary is a likely consequence.  This scenario  
was proposed as soon as the empirical limit was  recognized~\cite{hd79}.  
In a variant idea noted near the end of Section~\ref{3.1} below, 
the decisive mass loss occurs just before the stars become S Dor variables; 
but  both hypotheses invoke eruptions in that part of the HR diagram.    
No better alternative has appeared in the decades since they 
were~proposed.  

    \vspace{1mm} 

Today, S Dor variables are usually called LBV's,
``Luminous Blue Variables.''  Rightly or wrongly, they are frequently 
mentioned in connection with giant eruptions and supernova impostors. 
Many of them are easier to observe than eruptions in distant galaxies, 
and they probably offer hints to the relevant physics, Section~\ref{5}~below.   

   \vspace{1mm}  

Thus the key facts---episodic mass loss, and~the existence of 
giant eruptions---were well recognized before 1995, and~credible 
mechanisms had been noted; see many references in~\cite{hd94}.   
A decade later, when the mass loss rates of normal line-driven winds 
were revised downward~\cite{full06}, the~same concepts were 
proposed again as a way to rescue the published evolution tracks  
(e.g.,\ \cite{so06}).  Unrelated to that development, extragalactic 
giant eruptions attracted attention after 2003 and some of them were 
aptly called ``supernova impostors'' because their stars survived~\cite{vd05}.  Modern SN surveys 
found many examples~\cite{vd12},  often classed among the Type IIn 
SNae.   Some giant eruptions preceded real supernova events, the~
most notorious being SN 2009ip where the real SN explosion did not 
occur until 2012~\cite{past13,marg14}.  Theoretical explanations 
continue to be diverse and highly~speculative. 

      \vspace{2mm}  


\section{Categories and~Examples}\label{3}    
   \vspace{1mm}  

Only a few specific objects are mentioned here, to~illustrate 
the main~phenomena.   

\subsection{LBV's}\label{3.1} 

   \vspace{1mm} 

The term ``Luminous Blue Variable''  is unfortunate in three 
respects:   Many unrelated luminous blue stars are also variable, 
LBV's are often not very blue or not strongly variable, and  
the term has caused extraneous objects to be included in lists 
of examples, often without observed outbursts~\cite{rmh16,kd16}.          
Thus we should regard the trigram ``LBV'' as an abstract 
label, not an acronym.   Many examples are described and listed 
in~\cite{wb20,jv12,rmh16}.  The~present review includes this  
phenomenon because it may provide some guidance to the physics of 
giant eruptions, and~LBV's have been observed far more often 
than giant eruptions.  Note, however, that suspected analogies 
between those two categories have not been proven and may turn 
out to be~illusory.  

   \vspace{1mm}  

LBV's are defined by a particular form of variability like 
AG Car in Figure~\ref{f1}~\cite{hd94}.  Their hot ``quiescent'' states 
are located along a strip in the H-R Diagram (HRD) shown in the   
figure, sometimes called the S Doradus instability strip~\cite{bw89}.   
Its upper and lower parts, LBV1 and LBV2, represent different 
stages of evolution---providing a clue for theory, Section~\ref{5.1}~below.  

    \vspace{1mm}  

Most stars in the strip are not LBV's. Spectroscopic analyses consistently 
show that genuine LBV's have smaller masses than other stars with similar 
$L$ and $\teff$  \cite{jv12,kudr89,pp90,sta90,ste91,vdk02}.    
Consequently, for~an LBV the Eddington parameter 
$\Gamma \equiv {\kappa_e}L/4{\pi}cGM = (L/M)/(L/M)_\mathrm{Edd}$ 
is close  to 0.5 or somewhat larger.   This is not surprising for 
the luminous classical LBV's; a 60 $M_\odot$ star, for~example, 
attains $\Gamma > 0.4$ before the end of central hydrogen 
burning~\cite{rmh16}.   
But $\Gamma \sim 0.5$ is remarkable in zone LBV2, where most stars 
have $\Gamma \sim 0.2$.   Evidently each lower-luminosity LBV has 
lost much of its initial~mass. 

   \vspace{1mm}

Since the LBV2 stars have luminosities below the upper 
boundary in Figure~\ref{f1}, they can evolve across the HRD.  Hence we 
can explain their low masses by supposing that they have already 
passed through a cool supergiant stage where mass loss was   
very large~\cite{hd94}.  After~returning to the blue side  
of the HRD, they now have large $L/M$ ratios which cause them to 
be LBV's.  This surmise is very strongly 
supported by the fact that LBV1's are generally associated with 
O-type stars but LBV2's are not~\cite{rmh16,kd16,aad18}.  
Therefore LBV1's are younger than LBV2's, as expected 
in the evolved-LBV2 scenario. 
Classical LBV's (LBV1's) are somewhat more than 3 million 
years old, near~or slightly after the end of core hydrogen burning.  
LBV2 stars are post-RSG's near or after the end of core helium 
burning (Figure~1 in ref.\ \cite{rmh16}). 

   \vspace{1mm}  

The above account may seem inconsistent, because~LBV's are said 
to have rapid mass loss but the low masses of LBV2's are ascribed 
to a different evolutionary stage. This semi-paradox arises because 
the two types  play very different roles in this story.  LBV1 
outbursts probably cause enough mass loss to shape the appearance 
of the  upper H-R Diagram.  LBV2 events do not, but~they 
are pertinent because they suggest a connection between LBV 
variability and the Eddington parameter $\Gamma$ as noted above.  
They also give a strong hint that LBV instability occurs in the 
outer layers, Section~\ref{5.1}~below.

   \vspace{1mm}

Figure~\ref{f1} shows a well-known classical LBV, 
AG Carinae.  It currently has $L \approx 1.5 \times 10^6 \; L_\odot$ and 
$M \sim$ 40 to 70 $M_\odot$, with~$\teff \sim$ 16000 to 25000 K 
at times when a major LBV event is not underway~\cite{jv12,sta01,gro09,gro11}.
The~initial mass was probably above 85 $M_\odot$ and rotation 
is non-negligible~\cite{gro11}.  In~the years 1990--1994, AG Car's 
photosphere temporarily expanded  by an order of magnitude with 
only a modest change in luminosity~\cite{sta01}.  The~apparent 
temperature consequently 
declined to about 8500 K, shifting much of the luminosity to visual 
wavelengths.  Meanwhile its mass-loss rate increased by a factor of 
5 to 10, peaking above  $10^{-4} \; M_\odot$ y$^{-1}$.  (The estimated 
amount depends on assumptions about the wind's inhomogeneity.)  Outflow 
speeds varied in the range 100--300 km s$^{-1}$.  Then, in~1995--1999 
the photosphere contracted back to roughly twice its pre-1990 size.  
The event timescale, about 5 years, was more than $100 \times$  longer 
than the star's dynamical timescale.  Perhaps 5 years was   
a thermal timescale for a particular range of outer layers.    
AG Car's 1990--1999 event in Figure~\ref{f1} represents the classic form of 
high-luminosity LBV event, except~that it only partially returned 
to its pre-1990~state.  

   \vspace{1mm}  

Like many other LBV's~\cite{kw12}, AG Car has a circumstellar  
nebula~\cite{th50,no95,sm97,vn15}.  The~nebular mass is said to be 
5--20 $M_\odot$,  ejected thousands of years ago and expanding rather 
slowly.  Either the ejecta from multiple events  have piled up there, 
or the star had one or more giant eruptions larger than any LBV 
events that have been observed in recent times.   
(The circumstellar material is almost certainly not due to mass loss 
in a red supergiant stage of evolution, since AG Car is too luminous 
to become a RSG---see Figure~\ref{f1}.  The~same statement applies to 
various other LBV's that have circumstellar ejecta.)  

   \vspace{1mm} 

Figure~\ref{f1} includes another LBV, R 71, to~show that rules 
can be broken.  It had an outburst in the 1970's~\cite{was81}, 
but a later event starting around 2005 was   
extraordinary~\cite{meh13,meh17}.  Unlike normal LBV events, the~
luminosity of R 71 substantially increased while the temperature
fell definitely below 7000~K.  At~minimum temperature it exhibited  
pulsation on a dynamical timescale (cf.\ comments in~\cite{ji18a}).   
The mass-loss rate rose 
well above $10^{-4} \; M_\odot$ y$^{-1}$, high for its luminosity. 
Since $L$ is poorly known due to an uncertain amount of interstellar 
extinction, this object may be either a classical LBV1 or an~LBV2.  

   \vspace{1mm} 

Note that the empirical limit in Figure~\ref{f1} does not coincide with the 
LBV instability strip.  The~instability strip might extend to the 
boundary, but~this detail should warn us that evolution through the 
LBV1 stage may involve some unrecognized tricks.  Each of the following 
scenarios would be consistent with available~data.  
\begin{itemize} 
\item Standard LBV1 outbursts may cause enough mass loss to limit 
the star's later evolution.  However, the quoted rates and durations 
appear, at best, to be only marginally adequate.         
   \vspace{1mm}  
\item  Or perhaps the crucial mass loss occurs in rare, more 
extreme eruptions. P Cygni's dramatic brightening about 400 
years ago may have been an instance~\cite{pcygbook}, and~such an event 
may have created AG Car's massive ejecta nebula mentioned above.   
   \vspace{1mm}  
\item Conceivably the most important phenomenon occurs just {\it before\/}  
the LBV1 stage~\cite{sc93,gk93,sc94,sc99,lg14a}.  In~this scenario, the~
star first evolves across the LBV1 strip without incident, and~then 
becomes violently unstable at a stage near or beyond the empirical 
boundary.  A~giant eruption occurs, ejecting so much mass that the 
star moves back  to the left in the HR Diagram and becomes an LBV.  
The pre-LBV evolutionary episode would be too brief for us to have any 
known examples---though P Cyg and/or $\eta$ Car might 
fill that role (see below).  In~this view LBV's are results of the 
boundary, not its cause.  
\end{itemize} 
 
Resourceful theorists can devise other possibilities.  This review 
concerns the nature of mass-loss episodes, not the resulting 
evolutionary tracks.  The~latter depend on multiple parameters 
which are very poorly~known.     

   \vspace{1mm} 

Concerning LBV photospheres, see Section~\ref{4.5}~below.  

   \vspace{1mm}  

\subsection{Giant~Eruptions} \label{3.2} 

   \vspace{1mm}  

From a non-specialist's point of view, the observed giant eruptions 
have too much diversity. Some of them are supernova impostors (i.e., 
the star survives), while others may be genuine supernovae modified 
by surrounding  material (Section~\ref{5.4} below).  Both cases are 
often classified as ``Type IIn SNae,'' a term which implies 
narrower emission lines than normal SNae.  Major radiation-driven 
eruptions have several traits:   
\begin{itemize}
\item The flow is opaque during most of the event---i.e.,~the 
continuum photosphere is located in the outflow.  
\item Photospheric temperatures are usually in the range 6000--20,000 K 
defined in a particular way, Section~\ref{4.1} below.  
\item Outflow speeds are typically a few hundred km s$^{-1}$, not 
thousands, and~there are no conspicuous shock waves.  Small amounts of 
material may attain  higher speeds at the beginning of the eruption, 
but they are  relatively faint.    
\item H$\alpha$ and other bright emission lines have recognizable 
Thomson-scattered profiles as described in Section~\ref{4.2} below.  This fact 
is useful for indicating the nature of the eruption.  
\end{itemize}

   \vspace{1mm}

\begin{figure}[H]  
   \centering
   \includegraphics[width=10 cm]{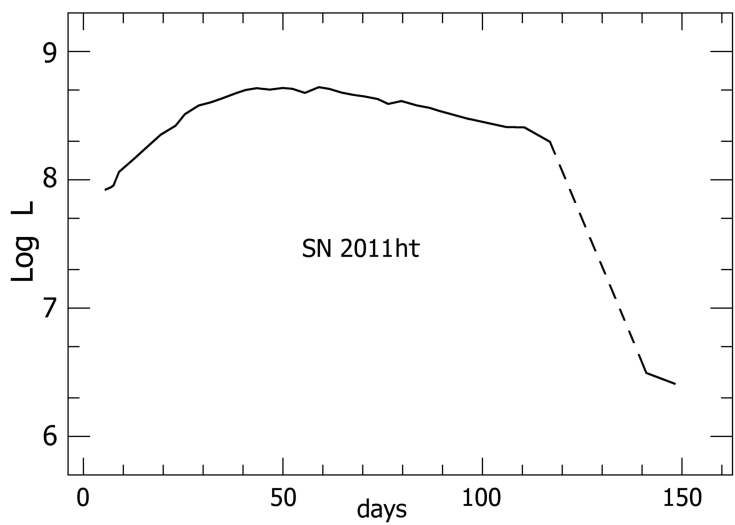}
   \caption{Luminosity record of SN2011ht based on visual-wavelength 
   brightness~\cite{rmh12}.   The~vertical scale,  expressed in solar 
   units, neglects variations in the bolometric correction but is adequate 
   for conceptual purposes.  The~relative faintness after $t \sim 130$ d 
   is highly abnormal if this object was a true supernova.}     \label{f2}     
   \end{figure}

An excellent example is SN2011ht~\cite{ro12,rmh12}, whose brightness 
and timescale resembled a supernova (Figure~\ref{f2}).  It may 
have been either a supernova impostor or else a true SN within a dense 
envelope of prior ejecta (Section~\ref{5.4} below); but in either case the 
observed display was a radiation-driven outflow.  Its spectrum  
(Figure~\ref{f3}) had characteristics explained in Section~\ref{4} below, with~outward  
speeds of several hundred km s$^{-1}$ and no hint of a blast wave 
before the brightness declined.  Broad emission line wings were 
caused by Thomson scattering rather than bulk motion (Section~\ref{4.2}),  
and the kinetic energy of visible ejecta was much smaller than in a 
normal SN~\cite{rmh12}.  

About two months after maximum, the~visual-wavelength brightness 
of SN 2011ht abruptly decreased by a factor of 60 (Figure~\ref{f2}).   
Since normal dust formation does not account for this change~\cite{rmh12}, 
the simplest interpretation is that most of the trapped radiation 
escaped through the photosphere just before that time. 
A normal core-collapse supernova would  
have remained substantially brighter due to radioactive decays in the 
ejecta.  Some authors {\it assumed\/} that 2011ht was a supernova in 
a discussion of the light curve~\cite{mau13};  but the lack of a 
radioactive afterglow was decidedly peculiar in that case, and~the 
light curve was reasonable for a non-SN instability (Section~\ref{5} below).   
The spectra in Figure~\ref{f3} gave far more definite information 
than the light curve did, and~strongly implied an opaque 
continuum-driven outflow 
far above the Eddington Limit~\cite{des09,rmh12,ka12}.  

   \begin{figure}[H]   
   \centering 
   \includegraphics[width=8cm]{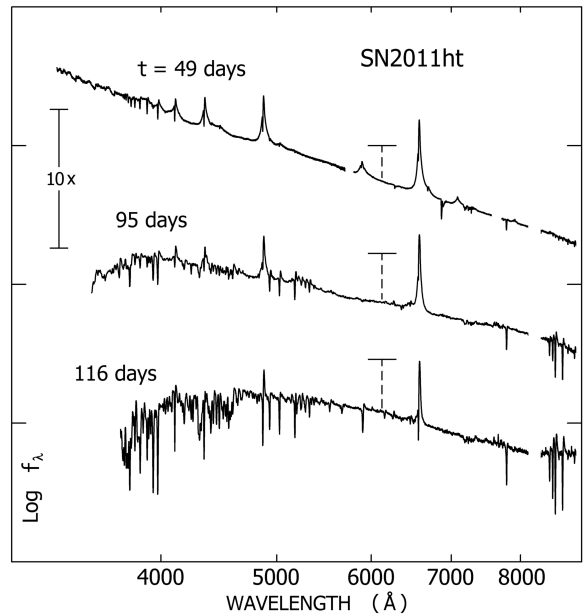}
   \caption{Spectrum of SN2011ht at three different times~\cite{rmh12}, 
   cf.\ Figure~\ref{f2}.  Both scales are logarithmic, the~three tracings have 
   differing vertical offsets, and~the marks near 6100 {\AA} indicate    
   $f_\lambda = 10^{-14}$ erg cm$^{-2}$ s$^{-1}$ {\AA}$^{-1}$.  
   Gaps at $\lambda > 7000$ {\AA} are obscured by terrestrial atmospheric  
   features.  Concerning the line profiles, see Section~\ref{4.2} and Figure~\ref{f6}. }  \label{f3}     
   \end{figure}

Historically, the~first observed giant eruption was P Cygni  about 400 years 
ago~\cite{wb20,pcygbook,rmh99}.   Its maximum luminosity was of the order of 
$10^{6.5} \; L_\odot$, quite small by giant eruption standards;  
but, unlike normal LBV events, that amount significantly exceeded 
the quiescent brightness.   P Cyg's outburst (actually two or more 
episodes)  persisted for years so the radiative energy output was 
probably more than $10^{48}$ ergs.  Today this star is located in 
the LBV instability strip in the HR Diagram, but~it has not exhibited 
an LBV event.  Possibly this is a hint that the interval between 
episodes is related to the strength of the most recent 
instance, analogous to some forms of relaxation~oscillators.

    \vspace{1mm} 

Among the eruptors discovered since 2000, some had luminosities 
comparable to $\eta$ Car and/or P Cyg, but~behaved very differently. 
SN2000ch, for~instance, has exhibited multiple outbursts considerably 
brighter than the familiar type of LBV outburst, but~not as bright as 
$\eta$ Car's Great Eruption~\cite{wag04,pas10,vd13}.  Those events were   
hotter than an LBV outburst, with~higher outflow speeds, much shorter 
durations, and~the luminosity may have increased appreciably on each 
occasion.  Altogether its behavior differs from LBV's, $\eta$ Car, and~
the brighter giant  eruptions noted below.  P Cyg might have appeared 
similar in the years 1600--1650, but~this is merely a~speculation. 

   \vspace{1mm}  

A more extreme object with a different kind of multiplicity was described  
in~\cite{hmgj16}.  PSN J09132750+7627410, a~SN impostor in NGC 2748,  
attained a luminosity of the order of $10^{7.3} \, L_\odot$, comparable 
to $\eta$ Car's maximum, for~several months---even though its quiescent  
luminosity was probably less than $10^{5.5} \, L_\odot$. 
Near maximum its spectrum resembled SN2011ht described above. 
Its chief peculiarity was the existence of several distinct outflow 
velocities in each absorption feature: $-400$, $-1100$, and~$-1600$ 
km s$^{-1}$.  These may signify either a series of mass-loss episodes,  
or structure in the observed episode, or~separate ejecta from more 
than one star.  The~two larger speeds are much faster than an LBV 
outflow.  Multiple velocities have been seen in a few other eruptive  
stars---e.g.,~see~\cite{marg14,mau15}.   

   \vspace{1mm} 

Two pre-2000 giant eruptions, SN1954J~\cite{ts68} and SN1961V~\cite{zw65,rmh99}, have had enough time to show whether their stars 
survived.    The~SN1954J event  
had a maximum luminosity of the order of $10^7 \, L_\odot$ 
with a duration less than a year~\cite{vd05,vd12,rmh17}.  The~
surviving star, a.k.a.\ V12  in NGC 2403, has a likely mass around 
20 $M_\odot$ and is seriously obscured by circumstellar dust. 
Its spectrum includes Thomson-scattered emission line profiles, 
indicating a present-day opaque outflow (Section~\ref{4.2} below).  This    
fact is strong evidence that the observed object really is the survivor 
of a giant eruption.  The~star was probably in a post-RSG state when 
the event occurred~\cite{rmh17}.  SN1961V, on~the other hand, remains 
doubtful.  It achieved a peak luminosity well above $10^8 \, L_\odot$ 
with an overall event duration longer than a normal supernova, but~
no survivor has been identified with high confidence~\cite{ko11,vdm12}.    

   \vspace{1mm} 

An unexpected development since 2000 has been the occurrence of 
precursor eruptions---i.e.,~giant eruptions that were followed 
several years later by real supernova events.   At~first sight 
this seems unlikely, because~the final stages of core evolution have 
timescales of days, hours, and~minutes rather than years.  A~few years 
is a likely timescale in the outer layers (Section~\ref{5} below), but~in the 
standard view those regions ``don't know'' the precise state of 
the core.  Hence one might conclude that precursor events arise 
in or near the core;  but that assessment is too glib to be entirely 
satisfying, as noted in Section~\ref{5} below.  The most notorious 
example of this phenomenon  
was SN 2009ip, whose  blast wave explosion was not observed until 2012~\cite{marg14}.  That object exhibited other events between 2009 and 2012.  
Evidently some part of the star became unstable a few years before the 
SN event, but~then the observed timescale didn't accelerate  
with the core evolution.  Or~perhaps the 2012 shock wave did not 
represent the real terminal event~\cite{past13,fra13a}!  See comments 
in Section~\ref{5.3}~below.  

   \vspace{1mm}  

SN 1994W, SN 2009kn, and~SN 2011ht probably ejected 
material months or years before their terminal explosions~\cite{des09,rmh12,fra13b}.  Since those objects became strangely faint 
at the stage when $^{56}$Ni decay normally produces luminosity after 
a core-collapse SN event,  some authors suspect that core collapse 
did not occur---e.g.,~\cite{des09}.  

   \vspace{1mm}  

Giant eruptions are usually easy to distinguish from 
LBV events.  They have far greater mass loss rates, substantial 
increases in luminosity, and~shorter durations in most cases.  
A few LBV's, however, have mistakenly been given SN designations.  
SN 2002kg, for~example, is a luminous LBV also known as V37 in 
NGC 2403~\cite{wb05,vd05,rmh17}.

   \vspace{1mm}  

\subsection{Eta~Carinae} \label{3.3}

    \vspace{1mm} 

The classic example of a supernova impostor, of~course, is $\eta$ Carinae.  
It merits a separate subsection here, because~it has been observed in 
far more detail than any other relevant object.    
Following the tradition of classic examples in astronomy, it 
has abnormal properties.   Its event seen in 1830-1860 persisted much 
longer than other known giant eruptions, and~it has a companion star 
that approaches rather closely at periastron.  Many authors 
reviewed $\eta$ Car in ref.\ \cite{dh12book}, and~later developments  
have not altered the 
main~facts.  

   \vspace{1mm} 

The star's luminosity is roughly $10^{6.6} \, L_\odot$ (Figure 1). 
Before 1830 its mass was probably in the range 140--200 $M_\odot$,  
with an apparent temperature close to 20,000--25,000 K~\cite{kd12}. 
Then its 30-year Great Eruption ejected 10-40 $M_\odot$ 
at speeds averaging 500 km s$^{-1}$. 
It converted more than $10^{50}$ ergs of energy   
to roughly equal portions of radiation, kinetic energy of ejecta, 
and potential energy of escape.  The~resulting ``Homunculus'' 
ejecta-nebula~\cite{ns12hom} is famously bipolar, indicating a 
complex role for angular momentum~\cite{kd12}.  The~ejected material 
is clearly CNO-processed, with~helium mass fraction $Y \sim$ 0.4--0.6~\cite{kd86,du99}. The~star's  subsequent recovery has been unsteady, 
including a smaller eruption around 1890 
and two later disturbances~\cite{kd12,hdk08,hm12}.  
Until recently its wind was opaque in the continuum with 
$\dot{M} \sim 10^{-3} \, M_\odot$ y$^{-1}$, and~most likely   
above $10^{-2} \, M_\odot$ y$^{-1}$ a century ago~\cite{hdk08}.  
That rate was too large for a line-driven stellar~wind.

The companion object is most likely an O4-type star with a highly 
eccentric 5.54-year orbit, see many articles in~\cite{dh12book}.  
The primary star is indeed the eruption survivor, since it has an 
extremely abnormal wind that has been diminishing since the event. 
But the hot secondary star alters the situation in several ways.  
\begin{itemize} 
\item  It ionizes much of the primary wind, greatly affecting the  
observed spectrum.
\item  Given the parameters and evolution of its orbit \cite{kd12,kd17}, 
it would have destabilized the orbit of any nearby third object 
thousands of years ago.  
\item Tidal friction near periastron can transfer orbital angular 
momentum to the primary star's outer layers.  If so, the 
equilibrium rotation period would be in the range 
50-150 days.  The primary star may be highly vulnerable to tidal 
effects because it is close to the Eddington Limit, but  
this possibility needs a careful quantitative analysis  
(see Section~\ref{6} below).      
\item Many authors have noted that the companion star may  
have triggered the Great Eruption via tidal influence near  
periastron~\cite{kd12};  but we must not confuse a trigger 
with the instability mechanism.  Perhaps the nearly-unstable 
primary star gradually expanded until the other star's tidal 
influence tipped it over the edge.  But~this idea is not 
simple, since there is  evidence for earlier episodes~\cite{dh97,kw12} and the present-day orbit eccentricity is too 
large to be caused entirely by the 19th-century mass loss~\cite{kd12,kd17}.  Incidentally, the~present-day orbital 
period cannot be used to estimate periastron times for the era of 
the Great Eruption. If we try to extrapolate back to about 1840, the
gradual period change due to fluctuating mass loss  
causes a phase uncertainty of the order of a~year.     
\item The companion star may be the main reason why $\eta$ Car's 
eruption was fainter and more protracted than most supernova  
impostors~\cite{kd12}.  For~several years the second star     
was inside the radius of the eruption photosphere, 
and near periastron it may have stirred the instability.    
During the great eruption, and~for many years afterward, 
the secondary star probably accreted some material from 
the primary's outflow~\cite{soker05,soker07,ks09,hdk08,kd12,kd15}.  
Possible consequences for the orbit have not yet been examined.  
\item As noted in Section~\ref{6} below, various authors have speculated 
that $\eta$ Car was originally a triple system and two of the
stars merged.  Models of that type have a large number of 
assumed parameters, they do not agree with each other, 
and there is no evident need for a third star.  
\end{itemize}

For historical reasons~\cite{hd79}, $\eta$ Car is often called an LBV 
despite its location in the HR Diagram (Figure~\ref{f1}).  The~high-luminosity 
end of the LBV strip in Figure~\ref{f1} may be misleading, and~in principle 
every star with  $L > 10^{6.3} \, L_\odot$ and $\teff <$ 25000 K might  
be an LBV; there are not enough examples to know.  But~that is only a 
possibility, and~we have no definite reason to classify $\eta$ Car 
as an LBV.  An~LBV-like eruption would be complicated for this object, 
because the radius of a 10000 K photosphere would exceed the companion 
star's periastron~distance.

\section{The Spectrum of an Opaque~Outflow}   \label{4}

This section has four main points: 
(1) Giant stellar eruptions usually have similar colors and spectra 
even if they're caused by different processes.   
(2) A particular type of emission line profile indicates an opaque outflow.   
(3) Stellar spectral types are not reliable indicators for outflow   
temperatures. 
(4) LBV outflows are not homologous with giant~eruptions.  

   \vspace{1mm}    

\subsection{The~Continuum }  \label{4.1}  

The apparent radiation temperature of an opaque outflow can be defined 
in various ways---e.g.,~based on the photon-energy distribution of the 
emergent  continuum, or~its slope at selected wavelengths, or~on 
subsets of absorption features, etc.  These alternative $T$'s can differ 
by 20\% or more, leading to confusion when one 
attempts to compare values quoted in papers.  ``Effective temperature'' 
$\teff$ used for stellar atmospheres is not appropriate, because~an 
outflow has no fundamental reference radius that is meaningful for 
that purpose.
Also note that an optical depth value of 2/3 has no significance 
in this context.  (Regarding photon escape probabilities, 
$\tautot \sim$ 1.0 to 1.3 in a diffuse spherical outflow corresponds to 
$\tautot \approx 2/3$ in a plane-parallel atmosphere.)  

   \vspace{1mm}  

And we must be careful with the word ``photosphere.'' The region with
optical depth $\tautot(r) \, \sim \, 1$ has little effect on an outflow's 
emergent photon energy distribution,  because~the dominant opacity 
is usually Thomson scattering by free electrons.  That process has only  
a weak effect on photon energies.  Consider instead a deeper 
region where absorption and re-emission events are frequent enough 
to establish $T_\mathrm{gas} \approx T_\mathrm{radiation}$.  
Outside some radius $\resc$,  the~average photon escapes via multiple 
scattering before it experiences an absorption event.  
Evidently the emergent photon energy distribution depends mainly on 
temperatures that exist just inside radius $\resc$.  In~this overview 
``photosphere'' means that~region.     

  \vspace{1mm} 

Classical diffusion theory gives the approximate size of $\resc$ 
\cite{rl79,kd87}.   
Suppose that local opacities for absorption, scattering, and~their sum 
are $\kabs$, $\ksc$, and~$\ktot$, averaged over photon energies in some 
optimal way.  Define a ``thermalization opacity''  
\begin{equation}    
     \kth(r)  \  \equiv  \ \left[ \, 3 \, \ktot(r) \, 
     \kabs(r) \, \right]^{1/2} \, ,    
   \label{eqnkth}
   \end{equation}
with associated optical depth    
\begin{equation}   
   \tauth(r) \ = \ \int_r^\infty \, \rho(r') \, \kth(r') \, dr' \, . 
   \label{eqntauth}
   \end{equation}
Often called thermalization depth or diffusion depth, $\tauth(r)$ 
is typically of the order of $0.6 \, \tautot(r)$ in a giant eruption 
or an LBV event photosphere.   Calculations show that $\resc$ 
is approximately the radius where $\tauth = 1$ \cite{rl79,kd87},  
and we can regard the photosphere as the region where  
$1  <   \tauth  <  2$.  This is not a formal statement, but~in 
practice it applies for any reasonable density law $\rho(r)$ and for large 
as  well as small opacity ratios $\kabs/\ksc$.  The~emergent continuum 
is created mostly at $\tauth \approx 1.5$ to 2.0,  while absorption 
and emission lines are formed mainly at $\tauth < 1$ or perhaps  
$\tauth < 1.5$.  If~$T_1$ and $T_2$ are the temperatures at 
$\tauth =$ 1 and 2, then we can liken $T_1$ to the 
$\teff$ of a star with a similar spectrum, though~their values  
may disagree because they are defined differently.  (Caveat:  
In published models of opaque winds,  some authors define the 
photospheric radius by $\tautot = 1$ or even $\tautot = 2/3$, 
rather than $\tauth = 1$.  With~those choices, a~quoted 
``photosphere temperature'' is cooler than 
the emergent distribution of photon~energies.)          

   \vspace{1mm} 

\begin{figure}[H]    
  \centering
  \includegraphics[width=11 cm]{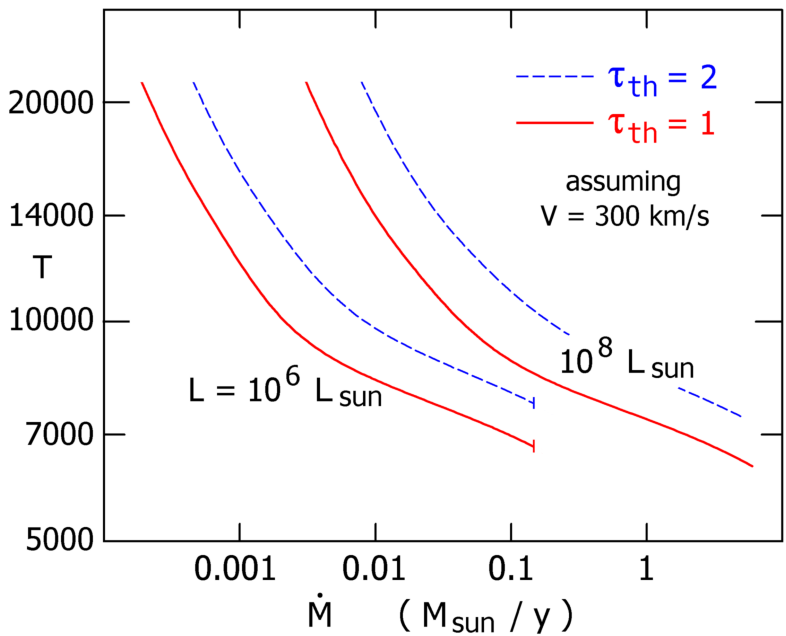}
  \caption{Photosphere temperatures in simplified opaque outflow models
   with $L = 10^6$ and $10^8 \, L_\odot$ and $V = 300$ km s$^{-1}$.
   The curves show temperatures corresponding to thermalization
   depths of 1 and 2.  $T(\tauth)$ depends approximately on the
   quantity $\dot{M} V^{-1} L^{-0.67}$.  Temperature values here are
   imprecise and very likely overestimated, because~the models are
   highly idealized; see text.}\label{f4}
  \end{figure}

\begin{figure}[H]   
  \centering
  \includegraphics[width=11 cm]{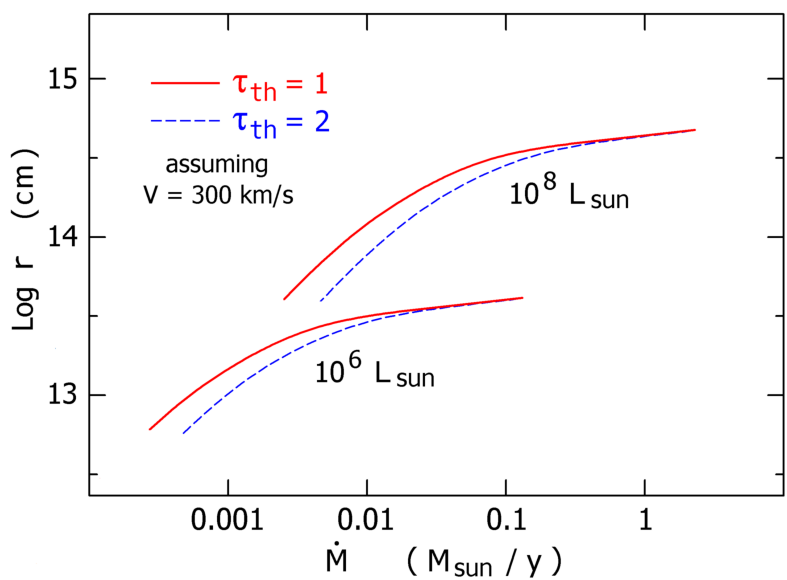}
  \caption{Radii of the photospheric locations shown in Figure~\ref{f4}.  For~large
   $\dot{M}$ the photosphere becomes geometrically thin (small $\Delta r/r$) 
   and thus resembles a plane-parallel model.}  \label{f5}
  \end{figure}

In a simple model where opacity depends only on $\rho$ and $T$,  
the  temperature at a given location depends approximately on two 
quantities, $\tauth$ and  $\dot{M} V^{-1} L^{-0.67}$ where $V$ is 
the local outflow velocity~\cite{kd87,os16}.  Figure~\ref{f4} shows examples 
of $T_1$ and $T_2$ in spherical outflows. Corresponding radii  
are shown in Figure~\ref{f5}.   These sketches are intended only for 
conceptual purposes;  they are based on simplified models that 
ignore some major details (see below).     

   \vspace{1mm}   

   \vspace{1mm}

With Figure~\ref{f4} in view, imagine a case with constant luminosity $L$ 
while $\dot{M}/V$ gradually increases.  Initially the flow is transparent   
so our model does not apply.  But~when $\dot{M}/V$ becomes large    
enough to be opaque, then it determines the apparent temperature.  
Increasing $\dot{M}/V$ causes the photosphere to move outward, and~  
$T_1 \, \propto$ roughly $(\dot{M}/V)^{-0.3}$ as shown in the upper 
half of Figure~\ref{f4}.  Below~9000 K, however, the~proportionality changes 
to $(\dot{M}/V)^{-0.07}$ because the opacities decline rapidly.  
$T_1 < 7000$ K requires a very large flow density.  As~noted above, 
$T_1$ is a fair indicator of the absorption and emission lines 
-- except for a caveat in Section~\ref{4.3}~below.   


  \vspace{1mm}  
  
$\bigtriangledown$ 
(This paragraph concerns technicalities that don't affect the 
main concepts.)  Each temperature in Figure~\ref{f4} refers to a location 
in the flow, which does not represent 
any specific observable quantity.  For~comparison with observations, 
one would need to calculate emergent radiation in a manner resembling~\cite{kd87} and~\cite{hr71};  but a model with wavelength-dependent 
opacities would be much better.  
Figure~\ref{f4} is based on many simplified models with constant 
luminosities $L$, mass-loss rates $\dot{M}$, and~flow velocities $V$.  
It was assumed that $T_\mathrm{gas}(r) = T_\mathrm{rad}(r)$, with~
LTE Rosseland mean opacities which are readily available   
(\url{http://cdsweb.u-strasbg.fr/topbase/}, \cite{opac05}, and~refs.\ 
therein).   Hydrogen and helium mass fractions were 
$X = 0.50$ and $Y = 0.48$.   Integrating the 
spherical radiative transfer equations~\cite{hr71} 
with those  opacities,  we can calculate $T(\tauth)$.  
One recent analysis~\cite{os16} appears at first sight to favor 
lower temperatures than Figure~\ref{f4}, but~the differences involve  
merely the distinction between $\tauth$ and $\tautot$, and~varying definitions
of the observed $T$.  True ionization in the outer regions tends to 
be larger than the LTE values;   if so then the temperatures    
in Figure~\ref{f4} are overestimates.  Errors of that type are probably 
comparable to the differences between alternative definitions of $T$ 
for the emergent~radiation.  

  \vspace{1mm} 

Figure~\ref{f4} is fairly consistent  with observed giant eruptions, such 
as $\eta$ Car marked in Figure~\ref{f1}.  Exceptionally high flow 
densities occurred in $\eta$ Car's 1830--1860 Great Eruption, with~
$\dot{M} > 1 \, M_\odot$ y$^{-1}$  and  $L \sim 3 \times 10^7 \, L_\odot$ 
\cite{dh12book}; so it should have had exceptionally low photospheric 
temperatures.   Reflected light echos provide spectra of that event~\cite{rest12,pri14,ns18a}.  The~light-echo researchers deduced a 
temperature near 5000 K, 
but that was an informal value, not well-defined and not based 
on a quantitative analysis.  Judging from the published description 
and reasons noted in Section~\ref{4.3} below, the~spectrum most likely indicated
$T_1 \sim$ 5500 to 6500 K in Figure~4~\cite{dh12nature}.  (Recall that   
$T_1$ denotes the temperature at a particular location in the outflow, 
not an emergent radiation temperature.) 
If we define ``apparent temperature'' in a different way, its value 
may have been as low as 5000 K~\cite{os16}.  A~second, lesser eruption 
of $\eta$ Car in the 1890's had $T_1 \sim 7500$ K, according to 
the earliest  spectrogram of this object~\cite{hdk08}.   
Several extragalactic giant  eruptions have been observed since 2000, 
usually showing $T_1 > 8000$ K (Section~\ref{3.2} above).  LBV outbursts 
have much smaller luminosities, and~usually reach 8000-9000 K like 
AG Car in Figure~\ref{f1}.  The~assumptions in Figure~\ref{f4} are not valid for them, 
because their photospheres are not located in the high-speed part of 
the flow (Section~\ref{4.5} below).  However, those observed minimum temperatures 
of LBV's are very likely determined by the rapid decline of opacity 
below 9000 K, even though the photospheres resemble static~atmospheres. 

  \vspace{1mm} 

In summary, there is no known observational reason to doubt the general   
appearance of Figure~\ref{f4} for opaque radiation-driven~outflows.  

   \vspace{1mm}   

\subsection{Distinctive Emission Line~Profiles}   \label{4.2}

\begin{figure}[H]   
  \centering
  \includegraphics[width=11cm]{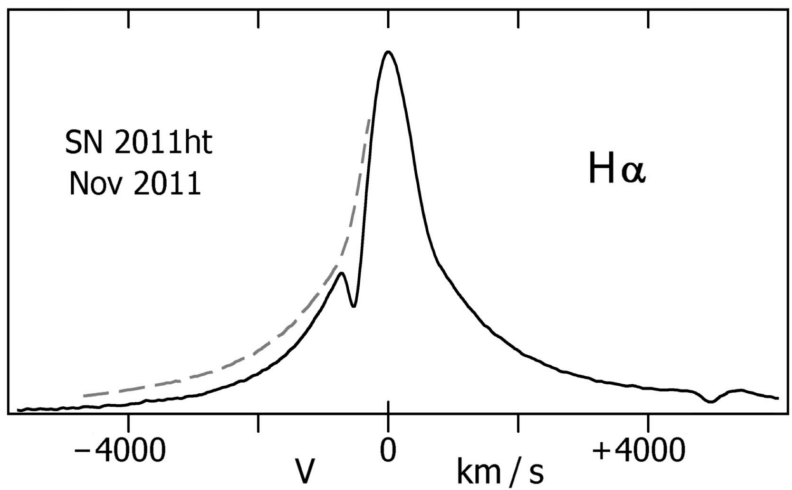}
  \caption{ Emission line profile with moderate Thomson scattering.  The~
  dashed curve on the left side is a mirror image of the right side.  
  Note that the line wings extend far beyond the velocity indicated by 
  P Cyg absorption.  This example is H$\alpha$ in the radiation-driven 
  outflow of SN 2011ht~\cite{rmh12},  but~other giant eruptions produce 
  similar line shapes.}\label{f6}
  \end{figure}

The brightest emission lines from an opaque outflow generally have a 
certain type of profile, illustrated in Figure~\ref{f6}. Smooth broad 
line wings extend beyond ${\pm}2000$ km s$^{-1}$ even though the wind 
speed is less than 700 km s$^{-1}$;  and the longer-wavelength side 
is stronger.  These are classic signs of Thomson scattering by free 
electrons~\cite{rw70,fos95,chug01,des09,rmh12}.  Since the electrons 
have r.m.s.\ thermal speeds of the order of 600 km s$^{-1}$, some 
of the photons acquire large Doppler shifts in multiple 
scattering events before they escape.  Meanwhile, expansion of the 
outflowing material favors shifts toward longer wavelengths.  Obviously 
the resulting profile depends on $\avtausc$, 
the line-emitting region's average optical depth for Thomson 
scattering.  The~shape in Figure~\ref{f6} indicates 
$\avtausc \sim$ 0.5 to 2, and~appears to be generic.  It specifically 
represents  SN 2011ht~\cite{rmh12},  but~SN 1994w exhibited a 
similar H$\alpha$ profile, and~so do other giant eruptions and 
$\eta$ Car's dense wind~\cite{des09,hm12,am15}. 

   \vspace{1mm} 

   \vspace{1mm}  

The moderate size of $\avtausc$ has a simple explanation.  In typical  
eruptions with $T_1 > 7500$~K, $\ksc/\kth \sim$ 1 to 2 ---a 
consequence of atomic physics.  Hence the continuum photosphere   
boundary $\tauth \approx 1$ automatically has $\tausc \sim$ 1 to 2. 
Since the emission lines are formed outside the photosphere,  we 
therefore expect $\avtausc \sim 1$, or~perhaps a little smaller, 
for bright emission lines.   The~main point here is that 
an observed profile like Figure~\ref{f6} is good evidence for an opaque 
or semi-opaque outflow.  It recognizably differs from emission 
lines seen in normal stellar winds, expanding shells, nebulae, 
and supernova~remnants. 

   \vspace{1mm} 

(Caveat:  In some papers a Thomson-scattered line profile is called 
``Lorentzian,'' often without recognizing its significance. 
That usage gives a flatly wrong impression in two respects.  
First, in~physics the word ``Lorentzian'' has very specific connotations: 
$1/(1 + x^2) =$ the Fourier transform of an exponential decay, 
the natural shape of an idealized spectral line, closely related 
to the uncertainty principle.  None of these applies to the shape 
in Figure~\ref{f6}.  Second, the~wings of a true Lorentz profile are like $x^{-2}$ 
but the wings of a Thomson-scattered profile are like $e^{-\alpha|x|}$.   
This difference is fundamental, not just a matter of opinion.)        

    \vspace{1mm}   

\subsection{Cautionary Remarks about Absorption~Features}  \label{4.3}

  \vspace{1mm} 

An opaque outflow also produces absorption lines,  but~they cannot safely 
be compared with stellar spectral types.  The~most dramatic example 
concerns $\eta$ Car's Great Eruption.  Spectra of that event have been 
obtained via light echos, leading to an estimate  $T \sim 5000$ K 
which seemed to contradict an expected value of 7500 K~\cite{rest12,pri14,ns18a}.  But~that conclusion had two disabling flaws.
(1) In fact the expected value  was far below 7500 K~\cite{kd87,dh12nature,os16}.   
(2) More pertinent here, spectral classification standards for stars do not apply 
to  a mass outflow.  For~instance the light-echo spectra of $\eta$ Car's 
Great Eruption showed absorption features of CN, which would indicate 
$\teff < 5000$ K in a star---but they may occur in an outflow 
with $T_1 \sim 6000$~K.     

  \vspace{1mm}  

\begin{figure}[H]
\centering
\includegraphics[width=9 cm]{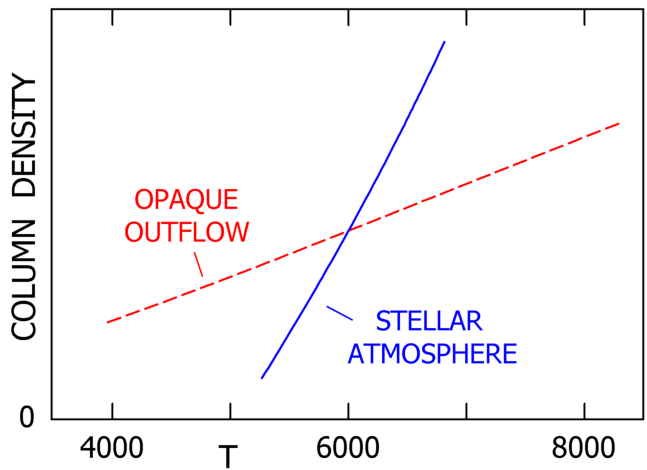}
\caption{Sketch of the mass column density at temperatures below $T$, in~
   a stellar atmosphere and in a dense outflow.  These curves are
      based merely on idealized textbook-style models of $\rho(r)$ and
         $T(r)$, but~the difference between them is qualitatively valid.}   \label{f7}
\end{figure}

Figure~\ref{f7} shows why.  Each curve represents the column density 
$\int \rho \, dr$ of material cooler than $T$.  A~stellar amosphere 
with $\teff \approx 6000$ K  has almost no material below  
5000 K, but~a diffuse outflow with $T_1 \approx 6000$ K can have 
an appreciable amount of cooler gas at large radii.  This difference 
is a consequence of two facts: (1) the mass distribution in an outflow 
resembles a power law $\rho(r) \propto r^{-n}$ instead of the exponential 
$\rho(z) \propto e^{-z/h}$ that roughly describes a stellar atmosphere, 
and (2) Radiation density in an outflow has a $1/r^2$ ``dilution factor.'' 
Figure~\ref{f7} is merely schematic,  but~it suggests that an eruption with  
$T_1 \sim 6000$ K can form cool spectral features such as CN in 
its outer regions.  Absorption lines formed at smaller radii may be 
good indicators of $T_1$, but~they must be chosen carefully.    
Since LTE is a poor approximation for $T < T_1$, and~there 
are other complications, a~realistic model of the absorption 
line spectrum will be extremely complex (see below).  Meanwhile, 
so far as available information allows us to judge, the~light-echo 
spectra of $\eta$ Car's eruption appear consistent with standard 
portrayals of that~event.   
   \vspace{1mm}  

   \vspace{1mm} 

\subsection{Why a Real Outflow Spectrum is Exceedingly  
Difficult to~Calculate}    \label{4.4}

The dependence of $T_1$ on $\dot{M}$ was first described long ago~\cite{kd87}.  Figure~\ref{f4} and ref.\ \cite{os16} employ modernized 
opacities, but~the resulting quantities are highly imprecise because 
the models are highly simplified.  A~truly realistic calculation 
must include all of the following~complications. 
\begin{enumerate}
\item Real opacities depend on photon energy, and~no form of average  
over $h\nu$ is fully consistent with all of the radiative transfer 
equations.  The~model should include $h\nu$-dependences. 
\item Standard LTE opacities are very unreliable in the region with 
$\tauth < 1.5$, because~it is far from thermodynamic equilibrium.  
Existing NLTE codes do not adequately include some effects, 
e.g.,\ items 4 and 5 below. 
\item A realistic model needs a good velocity law $V(r)$, and~
the functional form used for line-driven stellar winds is 
probably wrong for an opaque flow.  A~valid $V(r)$ is 
surprisingly difficult to calculate, because~of item 4 below~\cite{sh00,os12,gu18}.  This fact becomes even worse when we 
note that opacities due to Fe$^+$ and other complex species depend 
on many spectral lines whose interactions depend  
on item 4 as well as $dV/dr$!       
\item A radiation-driven outflow is obviously unstable---``a light 
fluid pushing a heavier one''---and breaks up into condensations, 
greatly complicating the radiative transfer problem~\cite{sh00,sh01,os12}.  Modern codes employ a position-dependent 
``clumping factor'' (e.g., \cite{gro11}) which entails several assumptions 
and free parameters that are very uncertain.  Hence the resulting 
models are useful guides but there is no reason to assume that 
they are correct.  If~each condensation is not transparent,  then 
radiation tends to escape via the easiest paths between 
condensations.  (This definitely happened in $\eta$ Car's giant   
eruption~\cite{sh00}.)   
The words ``porosity'' and ``granulation'' are often used in this context.  
Results depend on the condensation sizes, 
densities, and~even shapes.  A~practical technique, analogous to mixing 
length theory for convection, is needed for inhomogeneous radiative 
transfer. Perhaps a recipe can be developed from a large number of 
specialized three-dimensional simulations.  See~\cite{gu18,owo19}.
\item As $\eta$ Car notoriously shows, spherical symmetry may be 
a poor approximation.  Moreover, since  a spherical model 
maximally entraps the radiation, it represents an extreme case, 
not typical or average.  And~the spectrum of a non-spherical outflow 
depends on the observer's viewing direction.   
\end{enumerate}  
 
Omitting any of these complications may cause the results   
to be almost as inaccurate as the simplified models in Figure~\ref{f4}.    
Some existing codes employ elaborate radiative 
transfer with some NLTE effects, but~there are reasons for skepticism.  
Items 3, 4, and~5 have multiple undetermined free parameters, 
not advertised in most research papers.   Items 3 and 4 acting 
together may invalidate the radiative transfer methods for spectral 
lines. Most important, the~realism of a calculation is very 
difficult to test.  Approximately matching an observed spectrum does 
not prove correctness,   
since there are enough free parameters to compensate for omitted 
effects.  In~summary, existing codes illustrate most of the chief 
processes, but~they must not be regarded as decisive or authoritative. 
Their uncertainties may be far worse than most authors~assume.   

    \vspace{1mm} 

\subsection{Are LBV's Relevant?}  \label{4.5} 

    \vspace{1mm} 

As noted at the end of Section~\ref{3.1} above, the~most critical mass-loss 
episode may conceivably occur just before the LBV stage 
of evolution.  If so, then the relatively modest, familiar type of LBV 
event might have little in common with continuum-driven giant eruptions.  

    \vspace{1mm}   

The fast region of any LBV wind (usually 100 to 300 km s$^{-1}$) is 
transparent  in the visual-wavelength continuum. This fact was noted 
before 1990~\cite{kd87}, and~later motivated the static 
atmosphere models~\cite{lei89,dek96,gro11,gra12,sa15}.  But it doesn't 
necessarily apply to the inner, slow, denser part of the outflow  
during a major LBV event.  The~following remarks concern major LBV1 
eruptions with $T_1 <$ 10,000 K, not the states above 15,000 K 
that are emphasized in most analyses since~2010.    

    \vspace{1mm}  

Recall the analogy between a stellar wind and flow through a transonic 
nozzle, with~$V = w$, the~speed of sound, at~radius $r_w$ 
\cite{shbook,lcbook}.  The~subsonic region $r < r_w$ resembles an 
atmosphere constrained by gravity, while the flow dominates outside 
$r_w$.   Imagine such a hybrid model of AG Car's major eruption 
around 1994, when $T_1$ declined to about 9000 K while $\dot{M}$ 
rose to about $10^{-4} \, M_\odot$ y$^{-1}$ \cite{sta01}.   At~that 
time $w \sim 20$ km s$^{-1}$ in the photosphere, and the photospheric 
outflow speed was of the order of  $0.1 \, w$.   
The sonic  point was thus two or three scale heights above the 
photosphere,  much closer than in a normal line-driven stellar wind.  
According to Figure~\ref{f4}, a~$3\times$ larger 
value of $\dot{M}$ would have moved the photosphere to the sonic point    
-- so  a flow model rather than a static atmosphere would have become 
appropriate if $\dot{M}$ had grown above that amount.   In~this 
order-of-magnitude sense the major eruption was ``almost''~opaque.    

    \vspace{1mm}  

A normal line-driven stellar wind has an indirect relation to the 
continuum photosphere, with $V < 0.01\,w$ in the photosphere.   
The much larger 
$V/w$ in major LBV eruptions, along with proximity to the Eddington 
Limit, suggests that their outflows are more directly related to 
their continuum photospheres.  This line of thought---or rather, 
of surmise---motivates a three-step~conjecture:  
  \begin{itemize} 
  \item The rapid decline of opacity for $T < 8500$ K encourages the 
  eruptive photosphere to choose a temperature near that value. 
  In other words, for~a case with $L \sim 10^6 \, L_\odot$, the~
  basic parameters of the subphotospheric inflated region would  
  need to become unreasonable in order to push $T_1$ substantially 
  below 8000 K.  
  \item Due to the role of continuum radiation in the initial 
  acceleration, the~sonic point is related to the photosphere. 
  Consequently the outflow speed at $\tauth = 1$ is 
  $\, V_1 = {\xi} w \,$  where $\xi$ is of the order of 0.1.     
  \item Therefore, at times when $T_1 < 9000$ K,  
  $\dot{M} \approx 4 \pi r_1^2 \, \xi \, \rho_1 w 
    \, \approx \, r_1^2 \, \rho_1 w$ .    
  \end{itemize} 

This is an empirical hypothesis, not a theoretical 
prediction.  There are semi-theoretical methods of predicting  
LBV mass loss rates, far more elaborate than this reasoning, 
but their developed versions have been applied mainly to the hotter 
states above 15,000 K~\cite{jv18}.   Anyway, the~above expression   
is fairly consistent with estimated values for major LBV1~eruptions. 

   \vspace{1mm}   

If we portray LBV outbursts as quasi-static inflated states~\cite{lei89,dek96,gra12,sa15}, then unfortunately we de-emphasize   
the mass outflow.  Since the latter is probably more consequential, 
the ``eruption'' aspect expresses a broader significance than the 
inflation.   On~the other hand, it is conceivable that most of 
the cumulative LBV mass loss occurs in rare giant eruptions \`{a} la 
P Cygni, not merely major eruptions.  In~any case the static 1-D models 
do not answer the main questions, Section~\ref{5.1}~below.  

   \vspace{3mm}

\section{Physical Causes of the~Eruptions}  \label{5} 

    \vspace{1mm}

The Eddington Limit turns out to be wonderfully 
subtle and complicated.  Relevant instabilities were recognized 
after 1980 (see many refs.\ in~\cite{hd94}), but they're 
difficult to analyze or even to describe.  Practically all of 
our knowledge of eruption parameters, post-eruptive structure,
timescales, and~long-term mass loss is still~empirical.  

   \vspace{1mm}  

Two classic high-mass stellar instabilities were naturally suspected 
when giant eruptions and LBV events attracted notice in the 1980's: 
The Eddington $\epsilon$ mechanism or Ledoux-Schwarzschild instability 
energized in the stellar core~\cite{sh59}, and~radial dynamical instability 
which occurs if the average adiabatic index falls below 4/3~\cite{st99}.   
But they proved unsuitable~\cite{gl05}, and~have been replaced by 
newer ideas which are generally non-adiabatic, involve radiation
pressure, and~resemble each other.   Three regions in the star merit 
special attention: the photosphere,  the~``iron opacity bump'' region 
with $T \sim$ 200,000~K, and~the stellar core, broadly~defined.  

   \vspace{1mm} 

Details of the instabilities 
are too lengthy to explore here.  Instead, this overview lists a 
set of essential considerations, including some that have seldom 
been discussed in research papers.  Here rotation does not get the 
attention that it deserves, because~that would greatly lengthen 
the text.  Interacting or merging binary scenarios are omitted   
because there is no clear need for them at this time, see Section~\ref{6}.          

   \vspace{1mm} 

\subsection{The Modified Eddington~Limit} \label{5.1}  

   \vspace{1mm}  

Giant eruptions exceed the Eddington Limit.  Thus  
it seems natural to guess that the progenitor stars were very  
massive, close to the Limit, and~vulnerable to the same 
instabilities as LBV's.  This connection between LBV's and 
giant eruptions may be illusory, but~thinking about it leads  
to some useful~ideas.   

   \vspace{1mm}   

    \newpage

Observed $L/M$ ratios are clues to the LBV phenomenon in two  different 
ways.  First, of~course, their proximity to the Eddington Limit 
(Section~\ref{3.1} above) implies a large role for radiation pressure.  
Second and less obvious, the~distinction     
between two classes of LBV's favors an instability in the outer 
layers, not the stellar core.  Types LBV1 and LBV2 have 
different central structures because they represent different 
stages of evolution (Section~\ref{3.1} above and~\cite{rmh16}).  
The stars' outer 3\% of mass, however, have similar radiation/gas 
ratios for both classes, and~similar variations of~opacity.  

   \vspace{1mm} 

In the classical Eddington Limit
\begin{equation}
  \left(\frac{L}{M}\right)_\mathrm{Edd} \ \equiv \  
	  \frac{4 \pi c G}{\ksc} \; ,
  \end{equation}
the opacity $\ksc$ includes only Thomson scattering by free 
electrons.  Absorption opacity $\kabs$ practically vanishes as 
a static photosphere approaches the Limit, because~ density $\rho$ 
becomes very low.   The~classical Eddington parameter 
$\Gamma_\mathrm{Edd} \equiv \ksc L / 4 \pi c G M$  can thereby 
approach 1 in an old-fashioned radiative atmosphere model.  
However, the~photospheric $\kabs$  may be appreciable in a model with 
$\Gamma_\mathrm{Edd} < $ 0.95, and~deeper layers may have larger  
opacities in any case.  During~the 1980's this thought inspired 
the idea of a ``Modified Eddington Limit'' that takes $\kabs$ into 
account~\cite{hd84,ap86,la86,kd87x,ap89,hd94,asp98}.   It was 
an empirical hypothesis, not a theoretical prediction, motivated 
by $\eta$ Car and observed LBV~behavior.  

   \vspace{1mm}  

In principle there are two forms of Modified Eddington Limit. 
(1) It might be a well-defined limit to the allowed values of $L/M$ 
in a static stellar model, like the classical limit but including  
realistic $\kabs$, convection, incipient porosity, etc.   
(2) Or, more likely,      
it may signify an instability that arises when $L/M$ exceeds some 
value, see Section~5.2 in~\cite{hd94}.  In~either case the critical 
$L/M$ depends on opacities in the outer 3\% of the star's 
mass, or~maybe the outer 1\%.   

    \vspace{1mm} 

Rapid rotation reduces the effective gravity mass, thereby altering 
any form of Modified Eddington Limit.  The~terms ``$\Omega$ limit'' 
and/or ``$\Omega L$ limit'' allude to this obvious fact, 
but the implications  are often oversimplified.  
Two decidedly non-trivial subtleties occur:  
(1) Rotation causes a star's subsurface temperatures to depend 
on latitude.  Resulting alterations of opacity  affect the topics 
of Section~\ref{5.2} and \ref{5.3} below.  (2) The specific angular 
momentum expelled in a giant eruption may be either larger or 
smaller than in the lower layers.   Consequently the effects of 
rotation may evolve during the eruption~\cite{kd12}.  

    \vspace{1mm} 

\subsection{The Photosphere, Bistability, and~Surface~Activity}      \label{5.2}   

    \vspace{1mm} 

The easiest place to start is the photosphere.  Traditionally, the~
total energy flow in a stellar interior can exceed $4{\pi}cGM/\ktot$ 
by inciting convection~\cite{jso73}.  But~convection becomes 
inefficient in the photosphere, so radiation must carry nearly all 
of the energy flux there.  Imagine a static model wherein $\ktot$ 
increases inward, and~radiative forces are less than gravity 
outside some radius $r_c$.   Inside $r_c$, convection carries the 
excess energy flux.  But an increase in $L/M$  would presumably cause 
$r_c$ to move outward relative to the stellar material.  For~some value 
of $L/M$, $r_c$ moves into the photosphere;   so there may be a 
practical limit, somewhat smaller than $(L/M)_\mathrm{Edd}$.  By~
extrapolating normal atmosphere models,  one can identify a limiting 
$L/M$ around $0.9\,(L/M)_\mathrm{Edd}$ \cite{lf88,gp92,uf98}.  
But the reasoning is dangerously subtle, and~a different approach 
suggests that a star may become unstable at a smaller value of $L/M$, 
see~below. 

    \vspace{1mm} 

Next, instead of the above view, suppose that ``Modified Eddington Limit'' 
connotes an instability 
that arises somewhere in the range $0.5 < \Gamma < 0.9$, rather than 
a well-defined static limit.  At~relevant photospheric densities, 
$\ktot$  has a maximum in the vicinity of $T \sim 13000$ K, involving 
the ionization ratio Fe$^{++}$/Fe$^+$.  Thus a high-$\Gamma$ 
atmosphere may be very unstable in a particular range of $T$ around 
that maximum, and~might act as a relaxation oscillator jumping back 
and forth across the unstable range~\cite{ap86,ap89}.   

   \vspace{1mm}  

Behavior like that is observed at a somewhat higher temperature~\cite{pp90,lsl95,vdl99,jv12}.   Consider a standard hot line-driven 
wind model wherein the star gradually becomes cooler.   As~$\teff$ 
declines below  20000 K,  Fe$^{++}$ and other suitable ion species 
become numerous enough to drastically increase $\ktot$;  so the wind 
becomes slower and much denser.  The~transition occurs across a narrow 
range of $T$, hence the term ``bistability jump.''  It was noted 
around 1990 as a likely cause of LBV events~\cite{pp90,lsl95}.   
That idea originally meant a difference between two outflow states,
but it rapidly evolved into a bistability between two quasi-static 
states of the star's outer layers~\cite{lei89,dek96,gro11,gra12,sa15}. 
One state corresponds to a quiescent LBV, the~other occurs 
during an LBV eruption, and~intermediate states are more  
unstable due to the opacity maximum mentioned earlier.  
In the eruptive state, the~outer layers are greatly expanded      
or ``inflated.'' 

   \vspace{1mm}  

But a set of inflated and non-inflated models does not constitute a 
theory of LBV variability;  instead it plays a role more like an 
existence theorem in mathematics.  A~proper theory must address  
the following~questions. 
\begin{enumerate}
\item What is the state of the outer regions during a major event when  
$T_1 <$ 10000 K?  The most elaborate spectral analyses~\cite{gro11,gra12} 
focus instead on models with $T_1 >$ 15000 K, close to the bistability jump. 
The cooler state is more difficult but also more consequential.  Moreover, 
all 1-D models disallow some effects that are probably essential~\cite{ji18a}. 
\item Why and when does an LBV eruption end?  The star does not merely 
evolve into an inflated state and remain there until further evolution 
occurs.  Instead it jumps unpredictably back and forth between differing 
states.  Does a major LBV event  cease when a critical amount of mass 
or energy or angular momentum has been lost, or~are the reasons chaotic 
or related to inconspicuous  changes in the  stellar interior?  
\item Is the photospheric opacity behavior sufficient to cause  
an LBV event?  Or is the deeper iron opacity peak (Section~\ref{5.3} below)  
needed?    
\item The central LBV problem concerns mass loss, not the star's 
radius.  What factors determine the increased $\dot{M}\,$?  
Do they resemble the conjectures in Section~\ref{4.5} above?  Conventional 
line-driven wind theory is probably inadequate in this parameter 
regime (Section~\ref{4.4} above).  A~Monte Carlo radiative transfer technique  
predicts credible $\dot{M}$ values for LBV's in their hotter phases~\cite{jv18}, but~it omits many intricate effects seen in a 3-D 
simulation~\cite{ji18a}.  
\item What determines the event recurrence rate?  Is it like a 
relaxation oscillator wherein the recurrence time depends on 
details of the preceding event~\cite{ap86,ap89}, or~is there 
some form of periodicity?   P Cyg had an extremely large event   
400 years ago and has seemed quiet ever since~\cite{pcygbook}.   
\item What determines the timescale of a transition to the   
LBV-event state?   Is it a thermal timescale for some relevant 
set of outer layers? 
\item How large is the cumulative amount of mass loss?  Does it   
vary greatly or randomly among LBV's with a given luminosity?  
\item How strongly do these answers depend on rotation as well as 
chemical composition?   And how much do the LBV eruptions alter 
the surface rotation and composition?  
\item Do more extreme LBV eruptions occasionally occur, violent enough to 
substantially increase the luminosity while ejecting far more mass 
than usual?  Observed ejecta nebulae, e.g.,\ around AG Car, may be relics 
of such events.  They might account for most of the cumulative mass loss.   
\end{enumerate}

   \vspace{1mm} 

The last item pertains to giant eruptions. In~order to 
expel a mass which greatly exceeds that of the unstable region, the~
process  must be like a geyser: instability begins at the top and 
moves downward (relative to the material) until some factor stops it.  
In this way a photospheric instability might even cause a giant eruption.  
As outer layers depart, a~large reservoir of radiative 
energy is progressively uncovered.  At~any given time the configuration 
resembles a steady-state model, since the observed timescale is 
much longer than the dynamical timescale.  Presumably the eruption 
ends when conditions change at the base of the flow 
-- perhaps when it reaches some particular feature in the 
pre-eruption interior~structure. 

   \vspace{1mm} 

Another form of Modified Eddington Limit relates to dynamical 
processes rather than the temperature dependence of opacity.  
A static atmosphere dominated by radiation pressure    
tends to develop inhomogeneities, granulation, and~porosity 
like an outflow; see~\cite{sh01,os12} and many refs.\ therein.   
Resulting turbulence can engender MHD effects, even though the photosphere   
is well above the temperatures traditionally associated with stellar 
activity.   These phenomena may influence the outflow 
rate, and~might even determine it.  Conceivably, $\eta$ Car's 
dense wind a century ago~\cite{hdk08} may have involved stellar 
activity  analogous to a red supergiant!  Cf.\ \cite{ji15}.   
   \vspace{1mm} 

Most of the above possibilities are not mutually~exclusive.  

   \vspace{1mm}  

\subsection{The Iron Opacity~Peak} \label{5.3}

   \vspace{1mm}

The ``iron opacity peak'' locale in a star, described below, is probably  
crucial;  but its instabilities are too complex for simple analysis, 
math expressions, and~predictions.  A~decisive analysis will require 
numerous 3-D simulations which have not been feasible so~far.    

    \vspace{1mm} 

Absorption opacity has a dramatic maximum at temperatures around 180,000 K, 
for reasons concerning ionization stages of iron. 
In a typical LBV-like very massive star,  $\ktot > 2\,\ksc$ throughout  
a temperature range such as 100,000 to 300,000 K,  though~the actual   
limits depend on mass density. Vigorous convection occurs there 
because ${\ktot}L$ obviously exceeds $4{\pi}cGM$, if~$L$ signifies   
the total energy flow.  Such a region offers a zoo of instabilities,  
and dynamically it decouples the outer layers from 
the stellar interior.  Since the associated mass and energy greatly 
exceed the photosphere, this region is the most promising part 
of the star for eruption mechanisms.  Its usual name, the~iron opacity 
peak zone, might be confused with the iron peak of cosmic abundances;  
and ``iron opacity bump zone'' is both inelegant and cumbersome.  
For convenience we'll use an acronym here:   OPR = iron 
opacity peak region in the star.  Although~it usually occurs in the 
outer 1\% of the star's mass distribution, its spatial radius may be 
considerably smaller than the stellar radius $R$.   
A second opacity peak will also be mentioned, involving helium 
at lower~temperatures.  

    \vspace{1mm} 

The OPR mass and energy are difficult to estimate from  
observational data.  If~$\mu(T)$ is the mass column density of 
layers cooler than $T$, and~radiation pressure dominates, then  
$\mu \approx  P/g \sim  aT^4/3g$ so
\begin{equation}   
     m(T) \ \sim \ 4 \pi R_T^2 \, \mu(T) \ 
	\sim  \ \frac{4 \pi R_T^4 a T^4}{3 G M_*} \, , 
     \end{equation}
where $R_T$ is a radius that has temperature $\approx 0.8T$.  But~the 
$R_T^4$ factor is quite uncertain, because~$R_T $ may lie deep within 
an extended envelope.  Consider, for~instance,  $\eta$ Car before its 
giant eruption.  If~we know only that $M_* \approx 150 \, M_\odot$, 
$L \approx 4 \times 10^6 \, L_\odot$, and~$\teff \approx$ 20,000 to 
25,000 K~\cite{kd12}, then the mass in the temperature range 
100,000--300,000 K may have been anywhere in the range 0.002 to 0.1 
$M_\odot$.  The~thermal timescale for this OPR might have any value 
ranging from a few days to a few months, depending partly on how we define it.  
Given these strong dependences,  the~effects of OPR instabilities may 
be very sensitive to the evolutionary state and structure of the star---and thus consistent with observed facts about giant eruptions and 
LBV's.  

    \vspace{1mm} 

A standard LBV eruption might expel no more than the OPR  mass, and~
the two amounts may even be related.  But~a giant eruption rooted in 
that region  must be geyser-like (Section~\ref{5.2}).  Some forms of instability   
cannot easily function like geysers, for~reasons involving timescales 
-- see a remark later~below.      

   \vspace{1mm}  
  
   \newpage  

Since 1993, almost every stability analysis of very massive stars has 
emphasized ``strange modes'' of pulsation~\cite{gg90,gk93,sc93,gl05,saio09,guzlo14,lg14a,gu99}.      
Apart from mathematical details, they have the following~attributes.    
  \begin{enumerate}
  \item  Strange modes are essentially dynamical rather than thermal.   
  They resemble accoustic waves, in~contrast to thermodynamic Carnot-cycle 
  pulsations driven by the $\kappa$ mechanism in lower-mass stars. 
  \item  Hence they are fundamentally non-adiabatic. They become especially 
  strong if the local thermal timescale is shorter than the dynamical 
  timescale.  
  \item  They occur if radiation pressure exceeds gas pressure. 
  \item The density dependence of opacity, ${\partial}\kappa/{\partial}\rho$, 
  is more critical than ${\partial}\kappa/{\partial}T$.     
  \item Purely radial strange modes can occur, but~non-radial modes 
  may be more important.  
  \end{enumerate} 
These characteristics are almost perfectly suited to the OPR  
in a star near the Eddington Limit.  Item 3 causes the local 
mass density to be relatively low, thereby enabling item 2.  
For a very brief account of strange modes, see~\cite{saio09}. 

   \vspace{1mm} 

Altogether, then, in~a star near the Eddington Limit, the~OPR 
forms a queasy sort of cavity between the stellar interior and the 
outer layers---with strong consequences for pulsation modes.  Even 
if we consider only 1-D radial motions, gas-dynamical simulations 
reveal phenomena that appear crucial for LBV's and giant eruptions~\cite{lg14a,gu18,guzlo14,lg14b,og05}.  An~essential factor is the 
time dependence of convection.  Normally a massive stellar interior 
obeys the Eddington Limit by shifting some of the energy flux to 
convection where necessary~\cite{jso73}.  
But this assumption fails in a structure that changes rapidly, 
e.g.,\ in pulsating layers.  Convection needs some time to  
develop, and~the dominant convective cells have finite turnover 
times.  Hence the convective energy flux lags behind the total    
energy flux, especially in the circumstances listed above for 
strange modes.   As~explained in the papers cited above, this fact 
causes the radiative flux to exceed the Eddington Limit at some 
times and places in a pulsation cycle.  No actual runaway outburst  
occurred in the simulations, but~their boundary conditions and 
lack of non-radial modes tended to inhibit such a development.   

    \vspace{1mm}

Three-dimensional simulations show the spatial fluctuations 
of convection, and~reveal some opacity-related phenomena that cannot  
appear in the 1-D models \cite{ji15,ji18a,ji18b}.  For~instance,  
helium opacity can become large within clumps of gas that have been 
lifted to regions with $T <$ 70,000 K~\cite{ji18a}.  The~result is a second 
opacity-peak region,  indirectly caused by the iron opacity bump.    
Local regions in and below the photosphere can thus have large 
radiative accelerations.  The~outer layers become  supersonically 
turbulent, and~local parcels of mass can be ejected in a chaotic 
way.  In~this manner we begin to graduate from ``pulsations'' 
to ``stellar activity'' or even ``weather''---
see Figure~2 in ref.\ \cite{ji18a}.    Unfortunately, the~3-D 
calculations are so expensive in CPU time that only a few have 
been~attempted.  

   \vspace{1mm}  

Given the facts outlined above, the~OPR is very likely the 
root of the LBV phenomenon.  It is especially dramatic in stars with 
LBV-like $L/M$ ratios, and~it is rich in phenomena that appear relevant 
to the questions in Section~\ref{5.2} above.  Moreover, effects found in numerical 
simulations can help to accelerate the ejecta.  Therefore, contrary to 
most papers in this topic, we should not assume that LBV outflows are 
merely line-driven winds---especially during a major outburst  
(cf.\ \cite{quat16}).  

   \vspace{1mm} 

But can the OPR incite a giant eruption?  No simulation has yet 
produced an outright eruption.   Maybe this is so because the 
``weather''  analogy is apt!  A terrestrial atmosphere 
simulation would usually go for a long time before it produces 
a typhoon.  By~analogy, perhaps a stellar eruption results from an 
infrequent coincidence of several chaotic processes---a Perfect 
Storm.  Note that the inflated LBV model in ref.\ \cite{ji18a} was still 
expanding when the calculations ended after 700 dynamical timescales, 
only a few percent of a typical event~duration. 

   \vspace{1mm}  

As mentioned earlier, if~a giant eruption can originate in the OPR 
layers of the star, then it must be a geyser-style process with   
instability propagating downward  through the stellar layers---or 
rather, the~successive layers move outward past the instability zone.  
The energy budget thereby becomes complicated,  because~inner regions   
tend to contract in order to compensate for the lost  energy.  
As noted by~\cite{quat16}, the~resulting small increase  
in local temperature can increase nuclear reaction rates;  so the 
overall event may be indirectly powered by hydrogen burning.  Nearly 
all of the mass is close to dynamical equilibrium throughout this 
process, but~thermal equilibrium fails in the outer regions.  This  
story may lend itself to additional instabilities deep within the 
star.  

    \vspace{1mm} 

Unfortunately the geyser analogy may fail for some types of OPR 
pulsational instability~\cite{gu99,gu05}.  When a pulse of material 
has been expelled, the~driving mechanism needs time to re-establish 
itself, and~that time may be much longer than the dynamical 
timescale.  In~that case the instability cannot easily 
propagate through deeper~layers.  

    \vspace{1mm} 

At first sight, a~supernova precursor eruption (Section~\ref{3.2}) cannot originate 
in the OPR,  because~such events happen only a few years before core 
collapse, and~the outer layers evolve much slower than that. 
In the outer layers, there is nothing special about the core's 
last few years.
But this view may be too naive, for~reasons noted in~\cite{arns14}.   
During those final years, turbulence in the core can generate  
unsteady burning and outward waves, which tend to expand the outer  
layers---``an early warning system for core collapse.''     
The OPR is so sensitive that it may respond violently to even a 
small change in the outer-layer structure.  Thus it seems conceivable 
that the opacity peak might play a role in every class of eruption 
from  LBV events to pre-SN~outbursts. 

    \vspace{1mm} 

\subsection{Instabilities in and Near the Stellar~Core}\label{5.4} 

Some giant eruptions probably originate near the centers of massive stars, 
rather than in the OPR.  But~the definite examples concern true supernovae 
in special circumstances, and~the nature of SN impostors (i.e., giant 
eruptions that are not related to SN events) remains~murky.   

    \vspace{1mm} 

A supernova can produce a radiation-driven eruption instead of a 
visible blast wave.  Suppose that a star produces an opaque 
mass outflow in the years preceding its SN explosion.  In that case, 
after the SN blast wave emerges from the star it moves into the 
surrounding opaque ejecta.  If the circumstellar density is favorable, 
photons can then diffuse outward faster than the shock speed 
\cite{chug04,ci11,mor12}.  Radiation thus reaches the $\tau \sim 1$ 
radius substantially before the shock does;  indeed the shock may 
emerge long after the time of maximum light.  The visible event 
represents ``photon breakout'' rather than ``shock breakout.'' 
Maximum luminosity is far above the Eddington Limit.   

   \vspace{1mm}  

The photon diffusion rate can be described in terms of a random 
walk, but the familiar version of that concept doesn't give a unique 
diffusion speed for comparison with the SN shock speed.  Instead, 
for conceptual purposes, here's  a formal example with a well-defined 
constant diffusion speed.    
Consider pure scattering in  a spherical configuration;  
absorption and re-emission are equivalent to scattering so far 
as the total energy flux is concerned.  
Suppose that the scattering coefficient is  
$k(r) = \zeta / r$, with a constant parameter $\zeta$.   
(In the notation of Section \ref{4} above, $k = \rho \kappa$.)   
In this case 
the time-dependent diffusion equation has a similarity solution 
that represents an expanding pulse of radiation density:  
\begin{equation} 
    U(r,t) \ = \ \left(\frac{E}{8\pi}\right) \,   
	  \left(\frac{3\zeta}{ct}\right)^3 \,  
    \; {\exp}\left(-\frac{3\zeta \, r}{c \, t}\right) \;  ,     
  \end{equation} 
which has total energy $E$.  Because of the choice $k \propto r^{-1}$
(which is admittedly unrealistic), 
this expression contains a velocity-like ratio $r/t$.  
At any given location 
$r$, the~maximum radiation flux occurs at $t = 0.75 \zeta r / c$ when 
about 24\% of the energy has passed.  At~any given time, half of the 
radiation is located outside radius $r_{1/2} \approx 0.9 ct / \zeta$; 
so the median diffusion speed is approximately $0.9 c / \zeta$.  
About 10\% of the radiation energy moves outward faster than 
$1.8 c / \zeta$.  If~$\zeta$ is small enough for this speed to outrun 
the SN blast wave, but~large enough to make the pre-SN outflow opaque 
-- say $1 < \zeta < 40$---then a radiation-driven eruption rapidly 
develops.    

    \vspace{1mm}  

In a more realistic case with $k(r) \propto r^{-2}$ rather than 
$r^{-1}$, the~diffusion speed accelerates outward.  
The light curve resembles Figure~\ref{f2}, with~
a sudden decline after most of the radiation has passed through 
the photosphere.  Meanwhile, of~course, the~radiation accelerates 
the mass outflow.  Later the SN blast wave may emerge after the 
brightness has declined, with~only a modest display.   Thus 
SN 2011ht,  for~instance, may have been either a true supernova 
with a hidden shock, or~an impostor with no shock~\cite{rmh12}.  

   \vspace{1mm}    

One point about shrouded supernovae is so obvious that it's often 
underemphasized:  {\it the required circumstellar material was 
probably  ejected in one or more giant eruptions\/} with 
$\dot{M} > 10^{-3} \, M_\odot$ y$^{-1}$, years or decades before 
the core-collapse events (Section~\ref{3.2} above).   Many researchers  assume  
that the pre-SN stars were  LBV's, because~LBV's are the best-advertised 
eruptors.   But~this surmise is not entirely consistent, because~
the deduced amount of ejecta usually surpasses the familiar type of 
major LBV eruption~\cite{rmh12}.  
A giant LBV event (Sections~\ref{3.1} and~\ref{5.2} above) would be needed---i.e.,~much stronger than any LBV outburst observed in the 
past few decades.  If~such large eruptions really do occur as part 
of the general LBV story, they must be very infrequent.  Thus we should 
be very surprised if several known SN events were closely preceded 
by random LBV episodes on that scale.  It seems far more likely 
that the pre-SN outbursts were somehow related to the imminent  
core collapse, i.e.,~related to the core structure. Hence the 
deduced  pre-SN mass ejection probably had nothing to do with 
standard LBV behavior.  Those stars may have been LBV's, but~there 
is no good reason to assume that they were.  The~precursor events  
may have resembled the outbursts of SN 2009ip (Section~\ref{3.2} above),  
but with longer time~scales.   

   \vspace{1mm}

Pulsational pair instability attracted attention a decade ago 
with reference to supernova impostors~\cite{wbh07,heg12,chw12,lnb19}, 
because it can produce repeated eruptions.  A~star with initial 
mass around 150 $M_\odot$ eventually becomes a pair-production
supernova, wherein core temperatures rise high enough to produce 
a significant rate of $\gamma + \gamma \rightarrow e^- + e^+$.  
This conversion of thermal energy to rest mass causes a pressure deficit, 
while the adiabatic index falls well below 4/3 which implies dynamical 
instability.  Hence the core begins to collapse, raising the 
temperature so the pair creation accelerates, and~runaway nuclear 
reactions unbind the whole star.  But~if the star's mass is somewhat 
smaller, then the central region stabilizes before it is entirely 
disrupted, and~the episode can repeat.  This repetition motivates 
the term ``pulsational'' instability.  It must be very rare because 
it occurs only in near-terminal stages of very massive stars.  
The phenomenon seems too indeterminate to be really satisfying;  the 
time interval between events is extremely sensitive to obscure 
details, and~the first such event probably expels all the hydrogen. 
For the latter reason, supernova impostors such as $\eta$ Car  
presumably did not involve this type of event.  Apart from having 
too many syllables, the~main fault of pulsational pair instability 
is the difficulty of making definite statements about~it. 

   \vspace{1mm} 

Parallel to the computational developments noted in Section~\ref{5.3}, 
3-D simulations have revealed new phenomena 
in the star's core region. An~important 
fact is that some numerical techniques, especially in 1-D models, 
entail artificial (i.e., illusory) damping of fluctuations. 3-D  
convection and turbulence become particularly vigorous during 
a massive star's 
final years~\cite{arnm11,arns14}, with~dynamic effects that  
cannot be represented in 1-D calculations.   Turbulence generates 
gasdynamic waves, which carry energy outward.  Consequently the 
outer layers, feebly bound because they are close to the Eddington 
Limit, expand or perhaps even erupt.  Mass ejection may occur~\cite{qs12,sq14,arns14},  
while the turbulence also causes the 
nuclear burning to be unsteady or even explosive.  The~outer 
layers  are quite vulnerable because their binding energy is much 
smaller than the nuclear energy being processed in the central 
region.   As~mentioned earlier, the~opacity-peak region may   
produce enhanced instabilities because of the waves flowing 
through it.  Given these circumstances, perhaps we should not be 
surprised that paroxysms occur just before core~collapse.  

   \vspace{1mm} 

What can we say about core-based eruptions that are {\it not\/} 
related to a SN event?  The processes mentioned above would 
not be suitable.  Eta Carinae, for~instance, still has 
considerable hydrogen even after its Great Eruption.  
Evidently it has not yet evolved far enough to have an 
exotic core region.  It probably has a very capable opacity 
peak region, but~doubts about the geyser process (see above) 
may require a core-region instability instead.  One credible 
possibility has been suggested in refs.~\cite{gu99,gu05,guzlo14}.
In a very massive, moderately evolved star, gravity pulsation 
modes (like ocean waves rather than pressure waves) 
may become numerous and strong at the lower boundary of the  
region that still has some hydrogen.  Suppose that they grow 
enough to mix some hydrogen into the hot dense zones below 
that boundary.  The~resulting burst of hydrogen-burning would rapidly 
lift some material,  possibly ejecting a set of  
outer layers, and~then the remaining material would settle 
down.  Events of this type may recur on a thermal timescale, 
reasonable for an object like $\eta$ Car.  Some remarks in~\cite{quat16},  concerning enhanced reaction rates when 
a star's total energy has been reduced by mass ejection, 
may be relevant to this~idea. 

   \vspace{1mm} 

Explorations of core instabilities have naturally concentrated 
on the final pre-SN state, because~the structure is highly  
complex then and because SN-related processes are 
most fashionable.  With~the development of 3-D computation, 
however, unpredicted phenomena may appear at earlier stages  
of evolution;   anyway that's what we need for giant eruptions 
if the opacity-peak region turns out to be~inadequate.    

\section{Other~Issues} \label{6}  

   \vspace{1mm} 

This review has largely omitted stellar rotation despite its 
probable importance. Rotation would greatly lengthen the narrative, 
and would entail more free parameters. 
A traditional exploration strategy makes sense:  (1) Begin with simple 
non-rotating models, (2) learn whether the known processes can produce  
eruptions without rotation, and~then (3) explore the effects of 
angular momentum.  This topic has not yet reached stage 3.    
In view of the multiple parameters required for a distribution of 
angular momentum,  this approach is particularly justified for expensive 
3-D simulations (Section~\ref{5.3} above).  
Apart from $\eta$ Car as noted below~\cite{kd12} and the morphology 
of LBV ejecta-nebulae~\cite{kw12}, there is little observational 
evidence concerning angular momentum in radiation-driven~eruptions.    

     \vspace{1mm} 

The same attitude is even more justified for eruption scenarios that 
require interactions of binary or multiple stars, particularly merger 
events.  As~noted many years ago, speculations in that vein allow 
theorists to ``ascend into free-parameter heaven'' \cite{jsg89}.  
Generically they require either small orbits or unusual orbit 
parameters.   They are credible for lower-luminosity events 
that are not discussed in this review (e.g., red transients), because~
moderate-luminosity star systems are very numerous.  The~observed 
lower-luminosity outbursts can be explained by supposing that a tiny 
fraction of stars experience mergers and other exotic interactions.  
Stars with $L > 10^{5.5} \, L_\odot$, however, are scarce;  
so we should not see the observed number of LBV's and giant eruptors 
if unusual circumstances are required.  It is true 
that most massive stars have companions, but~only a small fraction 
of them are close enough for major interactions~\cite{km19}.  Equally 
important, {\it there is no evident need\/} for giant eruption models of 
that  type.  The~HRD upper limit in Figure~\ref{f1} applies to practically  
all stars above 50 $M_\odot$, not just those with close companions.  
The LBV instability strip becomes much harder to explain if we suppose 
that it depends on multi-parameter interacting binaries~\cite{hd94,kd16}.  
And, perhaps most important, the~single-star processes in 
Section~\ref{5} appear sufficiently promising until proven otherwise.  
In summary:  Binary and multiple-system 
phenomena certainly deserve attention,  but~they have not yet earned 
a well-defined place in the giant eruption~puzzle.   

    \vspace{1mm}  

Binarity does play a role for our best-observed supernova impostor, 
$\eta$ Car, but~it probably did not provide the basic instability 
mechanism. This object merits additional paragraphs here because  
its known abnormalities may offer some hints.     
For instance, consider the hot secondary star's
high orbital eccentricity, $\epsilon \approx 0.85$, with~a periastron 
distance only about 3$\times$ or 4$\times$ larger than the primary 
star's radius~\cite{kd17}.  Tidal effects are significant during about 
3\% of the 5.5-year orbital period, and~may have triggered the Great 
Eruption as noted in Section~\ref{3.3}.  But~this is not a straightforward 
idea!  When we take the Eddington factor $\Gamma$ into account, the~
companion star's maximum tidal effect is of the order of 
10\% as strong as effective 
gravity at the star's surface~\cite{kd12}.  The~iron opacity  
peak region is less perturbed  because it has a smaller radius, 
and the core region is practically unaffected.  Hence the periastron 
tidal-trigger conjecture requires an instability 
that began fairly near the surface---the geyser concept again.  
Moreover, the Great Eruption did not begin suddenly;  instead the star's 
brightness began to rise and fluctuate years earlier~\cite{frew04,hm12}. 
Later the mass outflow persisted long after tidal forces became negligible.  
Nonetheless the trigger concept has undeniable appeal.  One can easily 
imagine a star expanding due to evolution, until~it encountered a radius 
limit  enforced by its companion.  This differs from a familiar Roche 
lobe story in two respects:  it was close to the Eddington Limit, 
and~the tidal force made itself felt only for a few weeks near 
each~periastron.    Since that was comparable to the star's dynamical 
timescale, it was neither an adiabatic nor an impulsive perturbation.  

   \vspace{1mm}  

Two other points should be noted about $\eta$ Car's periastron passages.  
First, after~a sufficiently long time, tidal friction should cause 
the star's outer layers to rotate synchronously with the orbital rate  
at periastron, like the planet Mercury.  The~surface rotation period 
would then be roughly 90 days.  In~fact the X-rays show a quasi-period 
of that length~\cite{kd98}.  Second, why is the orbit so eccentric? 
Its period would be only about 130 days if it were circular with $r =$ the 
present-day periastron distance.   If~the orbit was circular a few 
thousand years ago, then the simplest explanation for large $\epsilon$  
has two or three parts: 
(1) Most of the eruptive mass loss must have occurred near periastron, 
in order to eccentrify the orbit. 
(2) Several giant eruptions like 1830--1860 were necessary in order 
to attain $\epsilon \approx 0.85$.  
(3) However, since that value is very high, some additional factor 
was probably needed---e.g.,~asymmetric mass flows.  See~\cite{kd12} 
and references~therein.  

   \vspace{1mm} 

Another oddity concerns $\eta$ Car's equatorial skirt of  
ejecta.  It is manifestly not a rotating disk, but~instead appears 
to consist of radial spikes \cite{dh12book,art11}.  
Velocities   and proper motions indicate that they formed at 
about the same time as the Homunculus~lobes.  If the radial features 
are ``merely'' illumination effects, then the asymmetries 
must exist closer to the star. 

   \vspace{1mm}  

Various authors have speculated that $\eta$ Car's giant eruption was 
a merger event, entailing a former third 
star~\cite{lp98,llpw98,kh03,pzvdh16,ns08}.  Their scenarios employ at least 
8 adjustable parameters, plus qualitative assumptions that are not 
emphasized, in~order to account for 5 or fewer observed quantities.    
There is no evident need to postulate  
a third object;  the primary star  appears well suited to the 
single-star ideas listed in Section~\ref{5} above.  (For instance, it 
is near the Eddington Limit without any reference to companion 
objects, and~probably has a substantial iron opacity peak region.) 
The most detailed merger model~\cite{pzvdh16} predicted too low 
a helium abundance, its stated quiescent brightness was far too 
low, and~it was vague about the ejecta morphology.  Exotic models 
offer considerable entertainment value, but so far there is no 
reason to suspect that  they are necessary.   The single-star 
processes in Section 6, modified by the known companion star, seem 
very promising for $\eta$ Car but have not yet been analyzed in 
much detail.  

   \vspace{1mm} 

High-velocity material associated with $\eta$ Car has been interpreted 
as evidence for either a blast wave or an merger event~\cite{ns18a,ns08}.  
Some outlying ejecta have Doppler velocities of 
1000--3000~km~s$^{-1}$~\cite{kw12}, and~light-echo spectra of the 
Great Eruption may show velocities as fast as 10,000 km s$^{-1}$ 
\cite{ns18a}.  However, other interpretations appear more likely 
according to the ``maximum simplicity'' criterion.  Judging 
from H$\alpha$ images of the outer ejecta, the~high-speed mass and 
kinetic energy are probably less than  $10^{-5} \, M_\odot$ 
and $10^{44}$ ergs, and~possibly much less.  These amounts are 
substantially smaller than the mass and thermal energy of the star's 
opacity peak region, for~instance.  If~an eruptive instability  
begins suddenly, a~small amount of leading material may be ejected 
at high speeds, reminiscent of the acceleration of a SN blast 
wave as it moves through a negative density gradient.  Indeed an 
acceleration feature like that can be seen in Figure~2 of~\cite{ji18a}.  
The standard super-Eddington flow becomes established after the initial 
transient burst.   This explanation may be wrong, but~it as well-developed 
as the exotic conjectures, and~more credible because it fits the 
other characteristics of $\eta$ Car's ejecta~\cite{dh12book}. 
Moreover, the~very-high-velocity line wings in the light echo spectra
are so faint that they may be either instrumental artifacts or 
features caused by Thomson scattering in small dense locales 
where $\tausc >> 1$.  

   \vspace{1mm} 

As emphasized in Section~\ref{4} above, the~brightness and spectrum of a radiation-driven  
eruption do not tell us much about the star and its structure.  However, the~
post-eruption behavior may imply some useful information.  At any given time 
during the event, the~entire configuration is close to dynamical equilbrium 
(including flow processes) but far from thermal and rotational equilibrium.   
This remains true after the event subsides, leaving a star with a peculiar 
thermal structure.  It should then recover---i.e.,~find a new equilibrium 
state---in a few thermal timescales.  This process has been observed 
in $\eta$ Car, 
and the record is interesting in two respects:  it has taken longer than the 
expected 50~years, and~it has been quite unsteady~\cite{kd12}.  Major changes 
occurred at 50-year intervals~\cite{hm12,kd12}, and~the spectrum has evolved 
more rapidly during the past  20 years~\cite{meh10,kd18}.  This temporal structure  
surely depends on the star's thermal and rotational structure.  A~preliminary  
assessment of the recovery problem was reported in~\cite{kashi16}, but~multiple  
3-D simulations are~needed.  

  \vspace{1mm}  

As mentioned near the end of Section~\ref{5.2}, stellar activity and turbulent 
MHD may occur in the outer layers of LBV's and/or related stars.  
This would not be terribly surprising, since one can write the 
Schwarzschild criterion in a form that looks much like the Eddington 
Limit.  The~point is that MHD waves, or~similar processes, may 
assist the outward acceleration forces, and~ might even produce 
violent~instabilities.  

   \vspace{1mm}  

Finally, a~point in Section~\ref{4.4} merits repetition because it affects this 
entire topic:  a radiation-driven outflow is difficult to calculate!   
If one writes 1-D analytic equations for radiative transfer and acceleration, 
they give nonsensical results because a real outflow automatically becomes 
inhomogeneous.  Acceleration and radiation leakage depend on the 
sizes, spacing, and~even the shapes of the granules.  These effects  
are too intricate to calculate ab~initio for every model or sub-model. 
Therefore it might be valuable, and~certainly would be interesting, 
to have some sort of general prescription based on many specialized 
3-D simulations.  As~a first step, those simulations could include 
only Thomson scattering.  What factors determine the characteristic size 
scales and time scales and density distributions?   
Cf.\ \cite{sh00,os12,quat16}.   

  \vspace{50pt}

\funding{This research received no external funding, and was supported primarily by photons.}   
\acknowledgments{I am grateful to R.M. Humphreys, J. Guzik, I. Appenzeller, C. de Jager, M. Schwarzschild, E.E. Salpeter, and A.S. Eddington for indicating good points of view for this topic.} 
\conflictsofinterest{The author declares no conflicts of interest.} 

      \newpage 

\reftitle{References}

\end{document}